\RequirePackage{fix-cm}
\documentclass[natbib,twocolumn]{svjour3}          
\smartqed  
\pdfoutput=1
\usepackage{graphicx}
\usepackage{mathptmx}      
\usepackage{latexsym}
\usepackage{lineno}
\usepackage{caption}
\usepackage{textcomp}
\usepackage{color}

\journalname{Climate Dynamics}
\begin{document}

\title{How to reduce long-term drift in present-day and deep-time simulations?}


\author{Maura Brunetti         \and  Christian V\'erard}


\institute{M. Brunetti \at
              GAP-Climate, Institute for Environmental Sciences, \\ 
              University of Geneva, \\ 
              66 Bd Carl-Vogt, 1205 Geneva, Switzerland \\
              Tel.: +41-22-379 06 25\\
              Fax: +41-22-379 07 44\\
              \email{maura.brunetti@unige.ch}           
           \and
           C. V\'erard \at
              Climatic Change and Climate Impacts Group, \\ 
              Institute for Environmental Sciences, University of Geneva, \\ 
              66 Bd Carl-Vogt, 1205 Geneva, Switzerland \\
}

\date{Received: date / Accepted: date}

\maketitle

\begin{abstract}
Climate models are often affected by long-term drift that is revealed by the evolution of global variables such as the ocean temperature or the surface air temperature. 
This spurious trend reduces the fidelity to initial conditions and has a great influence on the equilibrium climate after long simulation times. 
Useful insight on the nature of the climate drift can be obtained using two global metrics, {\it i.e.} the energy imbalance at the top of the atmosphere and at the 
ocean surface. The former is an indicator of the limitations within a given climate  model, at the level of both numerical implementation and physical parameterisations, while the latter is an indicator of the goodness of the tuning procedure. Using the MIT general circulation model, we construct different configurations with various degree of complexity ({\it i.e.} different 
parameterisations for the bulk cloud albedo, inclusion or not 
of friction heating, different bathymetry configurations) to which we apply the same tuning procedure in order to obtain control runs for fixed external forcing where the climate drift is minimised. We find that the interplay between tuning procedure and different configurations of the same climate model  provides crucial 
information on the stability of the control runs and on the  goodness of a given parameterisation. 
This approach is particularly relevant for constructing good-quality control runs of the geological past where 
huge uncertainties are found in both initial and boundary conditions.  We will focus on robust results that can be generally applied to 
other climate models.      
\end{abstract}
\section{Introduction}
\label{intro}

Paleoclimate simulations of the geological past are particularly challenging since initial conditions are not well constrained by sedimentary data, flawed by uncertainties in dating and spatial scarcity, while boundary conditions are often affected by large uncertainty in paleogeographic 
reconstructions~\citep[p.~86]{DeepPast2011}. 
As a consequence, deep-time simulations are not only limited by the usual biases of climate models, but also by additional biases which come from the construction of imperfect initial and boundary conditions. Moreover, the technique of restoring surface temperature and salinity to observed initial conditions,  which is used to improve the stability of coupled simulations in preliminary integrations~\citep{Sanchez-Gomez2016}, is not possible when accurate initial conditions are not available. In such a situation the tuning procedure assumes a crucial role \citep{2012JAMES...4.0A01M,2013GeoRL..40.2246G,HourdinEtAl2016}. In present-day simulations, the tuning procedure guarantees the construction of control runs, which are runs where the climate forcing ({\it e.g.}, from solar brightness, atmospheric concentration of greenhouse gases, \ldots) is held constant, with minimal spurious 
drift~\citep{Covey2006,SenGupta2012,SenGupta2013}. In paleoclimate simulations, the tuning procedure becomes even more important because it helps in constraining the possible values of parameters, thus reducing the number of climatic attractors that can be explored within a given 
climate model~\citep{2008Chaos..18c3121F}. 

In general, tuning is not well-documented in climate simulations but the scientific community is now more and more aware that tuning should be made a more explicit process and should be taken into account for evaluating and interpreting model results~\citep{HourdinEtAl2016}.  
Tuning is used to improve the performance of a model in reproducing a given climate. However, if tuning is disconnected from the development of 
improved physical parameterisations at the process level, the risk is to have heavily tuned models that mask the presence of systematic errors~\citep{Jakob2014}.  
We will present here a tuning procedure that highlights the goodness of a given parameterisation. The idea is to apply the same tuning procedure to a 
hierarchy of configurations of a climate model, characterised by a given complexity in the representation of its physical processes. We will take advantage of 
the modular design of the MIT general circulation model (MITgcm)~\citep{1997JGR...102.5753M,1997JGR...102.5733M,2004MWRv..132.2845A}, where, as in many other climate models, physical processes and parameterisations in each component of the climate system are designed as a module that can be activated at runtime, allowing one to change the complexity of the climate simulations by including/excluding different parts of the code.  
The result is that the link between tuning and process description narrows the size of the parameter space, with clear advantages in particular in the case of paleoclimate simulations.  

The tuning procedure can change from one climate code to the other~\citep{tuneFamous2011,gmd-6-1447-2013,HourdinEtAl2016}.
The optimal solution found by tuning only one parameter at a time can differ from the one found by perturbing multiple parameters systematically, using objective methods such as, for example, cost function optimisation (see \citet{HourdinEtAl2016} and references therein). 
Such objective methods are still not commonly implemented in climate groups nowadays, while 
in general tuning procedure is performed in two stages. In a first stage, the model components (atmosphere, ocean, land, sea ice) are finalised independently in forced mode. In a second stage, the components are coupled together and only few parameters are allowed to change (in order to avoid compensating errors). For example, in CESM (Community Earth System Model) these tuning parameters are 
the sea ice albedo, and the humidity threshold that controls the formation of low clouds \citep{2011JCli...24.4973G,2016Tang}.
We apply here the same tuning procedure to the case of MITgcm in coupled atmosphere-ocean-sea ice configurations. 
We consistently use the same procedure at each level of complexity of the model to obtain  quasi-equilibrium simulations over long time-scales. 
We start from considering present-day simulations with different physical parameterisations. This will allow us to better understand the limitations of the code and to set the robustness of the results. 
Then we move to deep-time paleoclimate simulations in order to set the right procedure for obtaining well-balanced control runs.      

In both present-day and deep-time simulations,  
the tuning parameters are optimised by checking the energy imbalance of the system under consideration at the top of the atmosphere (hereafter named TOA imbalance) and at the ocean surface, since the ocean becomes the dominant energy store in the Earth system on a timescale of about one year~\citep{PalmerMcNeall2014}.   
We show that useful information on the limitations of the activated modules and on the quality of the control runs 
can be obtained from these energy budgets.  

The paper is organised as follows: in section~\ref{section:method} we describe the coupled simulations and 
the suite of modelling set-ups. In section~\ref{section:results} we analyse the control runs for each configuration and we describe the method that links tuning and parameterisation. In section~\ref{section:conclusions} we discuss the relevance of the method for deep-time simulations and we draw conclusions.
 
\section{Model description and experiments}
\label{section:method}

The coupled model that we use for the present study is the MIT general circulation model~\citep{1997JGR...102.5753M,1997JGR...102.5733M,2004MWRv..132.2845A}. We have chosen this code since it is modular programmed, open-access and it can use cubed-sphere grids that turn out to be particularly useful when polar regions are covered by oceans, as was repeatedly the case in the geological past. 
In this code, the same dynamical kernel is employed for representing atmosphere and ocean 
dynamics~\citep{MarshallAdcroft2004} on the same cubed-sphere grid. 
The Gent and McWilliams scheme~\citep{1990JPO....20..150G} is used to parameterise mesoscale eddies, while the KPP scheme \citep{1994RvGeo..32..363L} accounts for vertical mixing processes in the ocean surface boundary layer and the interior. 
The 5-level SPEEDY physics package for the atmosphere, that is described in~\citet{Molteni2003}, comprises a four-band radiation scheme, boundary layer and moist convection sche\-mes, resolved baroclinic eddies and diagnostic clouds. Orbital forcing is prescribed at present-day values and the atmospheric CO$_2$ content is fixed to a constant value of 326 parts per million. The Winton thermodynamic model~\citep{2000JAtOT..17..525W} is used for the sea ice component. A 2-layer land model is also included~\citep{Hansen1983}. 

The first step in using a climate model is to construct the so-called {\it control run}, {\it i.e.} the quasi-equilibrium configuration 
obtained by setting the climate forcing to a constant level. 
The quasi-equilibrium configuration corresponds to a state where global metrics, such as the surface air 
temperature (SAT) or the global ocean temperature, 
reach stationary values so that the climate model does not experience drifts that prevent to study the system over long simulation-time. 
This quasi-equilibrium configuration is obtained after a spin-up phase. It has  been shown that   
long spin-up has the advantage of reducing the rate of climate drift~\citep{SenGupta2012}, but on the other hand, 
even in the presence of a small drift,  a long integration means that the climate state is more likely to diverge from the initial conditions~\citep{SenGupta2013}. 
From here the importance of constructing control runs with minimal drift from the beginning to obtain at the same time great fidelity to initial conditions and 
the possibility of performing long-term runs.  

Here we consider a low resolution cubed-sphere grid, where each face of the cube has $32\times 32$ cells, giving a horizontal resolution of ca.~2.8$^\circ$. The ocean has 15 vertical levels with different thickness, from 50 m near the surface to 690 m in the abyss. In this way, we can perform simulations over thousand years, which is the dynamical time-scale of the overturning in the entire ocean,  
in a reasonable amount of CPU time, allowing us to test for different parameters and configurations.    
One of the advantages of the MITgcm code is that the modules for a given parameterisation can be activated at runtime so that one can easily construct a hierarchy of models with different degrees of complexity by including or not a given physical process. The simulations considered in the present study 
according to our suite of modelling set-ups are listed in Table~\ref{tab:1}.  

{\tt Run1} is the reference run where, beside all the above listed parameterisations, we also included a modification of the bulk cloud albedo that has been implemented in 8-level SPEEDY~\citep{speedy8,2006ClDy...26...79K} in order to correct a too strong high-latitude solar radiation flux. 
This reference run has been chosen among a series of  numerical experiments 
where we changed the values of vertical diffusivity within the ocean and of snow/ice albedo 
(as listed in Table~\ref{tab:2}) so that the temperature drift in the 
global ocean temperature was minimal, following the tuning procedure used in other modelling groups~\citep{2011JCli...24.4973G,2016Tang}. These albedo values are within the observed range reported, for example, in~\citet{2011JGRC..116.4025N}.
Afterwards, only the relative humidity threshold $RH$ for low clouds (a parameter referred as {\it RHCL2} in MITgcm atmospheric module) has been changed for
tuning the suite of modelling set-ups and obtaining the corresponding control run.

In {\tt Run2} we consider a slight change in the input parameter $RH$ with respect to the value found for {\tt Run1} to show its impact on the energy budget. In {\tt Run3} we use a different parameterisation for the bulk cloud albedo (the one typically used in 5-level SPEEDY~\citep{Molteni2003}) in order to discuss the impact of a different parameterisation on energy budget and tuning procedure, and how to decide which parameterisation is 
better. In {\tt Run4} the kinetic energy dissipated within the ocean and the atmosphere is re-inserted into the system as thermodynamic energy allowing for an improved TOA budget. In {\tt Run5} we apply the same procedure used in present-day runs to a different deep-time configuration, namely the Callovian (165 Ma) bathymetry~(see \citet{BrunettiVerard2015} for an example of Callovian configuration).            

Our analysis is based on the calculation of annual means of 
globally integrated energy budget variables,  as suggested in~\citet{2016JCli...29.1639H} for a useful first-order  
diagnostic of the model behaviour. The TOA energy imbalance is calculated by summing the net shortwave radiation flux and the outgoing longwave radiation flux at each grid point, ($TSR - OLR$ in MITgcm atmospheric module, in units of [Wm$^{-2}$]) and then performing area-weighted averages. 
Positive values correspond to a net radiative warming of the planet. The budget at the ocean surface (in units of [Wm$^{-2}$]) is estimated in two ways: {\it (i)} from the net heat flux into the ocean ($TFLUX$ in MITgcm ocean component, positive if heat enters into the ocean surface and the ocean warms) and {\it (ii)} from taking the time derivative of the ocean heat content $H$ per unit area, defined as:
\begin{equation}
H(t) = \frac{\rho_o c_{p}}{A} \int T(x,y,z,t)~dx~dy~dz   
\end{equation}
where $\rho = 1023$~kg m$^{-3}$ is seawater density, $c_p = 4000$~J (K kg)$^{-1}$ is the specific heat capacity 
(same values used in~\cite{2016JCli...29.1639H}), $T$ is seawater potential temperature and $A$ is the ocean surface.  

The model is spun up from rest,  
without snow and ice covered oceans. Initial three-dimensional
distributions of potential temperature and
salinity are derived from the ocean climatological database (annual means)~\citep{Levitus_Burgett_Boyer_1994,Levitus_Boyer_1994}. Land mask, vegetation cover, soil albedo and runoff are given to initialise the coupled atmosphere-ocean model.  
Simulations are run until deep-ocean equilibrium and the last 100 yr are used for diagnostics.


\begin{table*}
\caption{List of simulations.}
\label{tab:1}
\centering
\begin{tabular}{lcccc}
\hline  
Name & Bathymetry & Description &  Cloud albedo as in & $RH$  \\
 &  &  &  $n$-level SPEEDY  & \\
\hline 
{\tt Run1} & Present-day &Reference run & $n=8$ &  0.8440  \\
{\tt Run2} & " & Sensitivity to RH & $n=8$ &  0.8500  \\
{\tt Run3} & " & Cloud albedo &  $n=5$ & 0.7677  \\
{\tt Run4} & " & Friction heating & $n=8$ &  0.7900 \\
{\tt Run5} &  Callovian & " &  $n=8$ & 0.9035 \\
\hline
\end{tabular}
\end{table*}

\section{Results} 
\label{section:results}

We describe here the results obtained from the analysis of the suite of modelling set-ups  
listed in Table~\ref{tab:1}. 

\subsection{Reference run: {\tt Run1}}
\label{sub:1}

The reference run, {\tt Run1}, is the result of the tuning procedure described in the previous section where the final tuning parameter is the relative humidity threshold $RH$ for low clouds. Thus, 
given the approximations of the set-up described in Section~\ref{section:method}, this control run can be considered as a good description of the pre-indu\-strial climate. 

\begin{figure}[t]
\includegraphics[width=9cm]{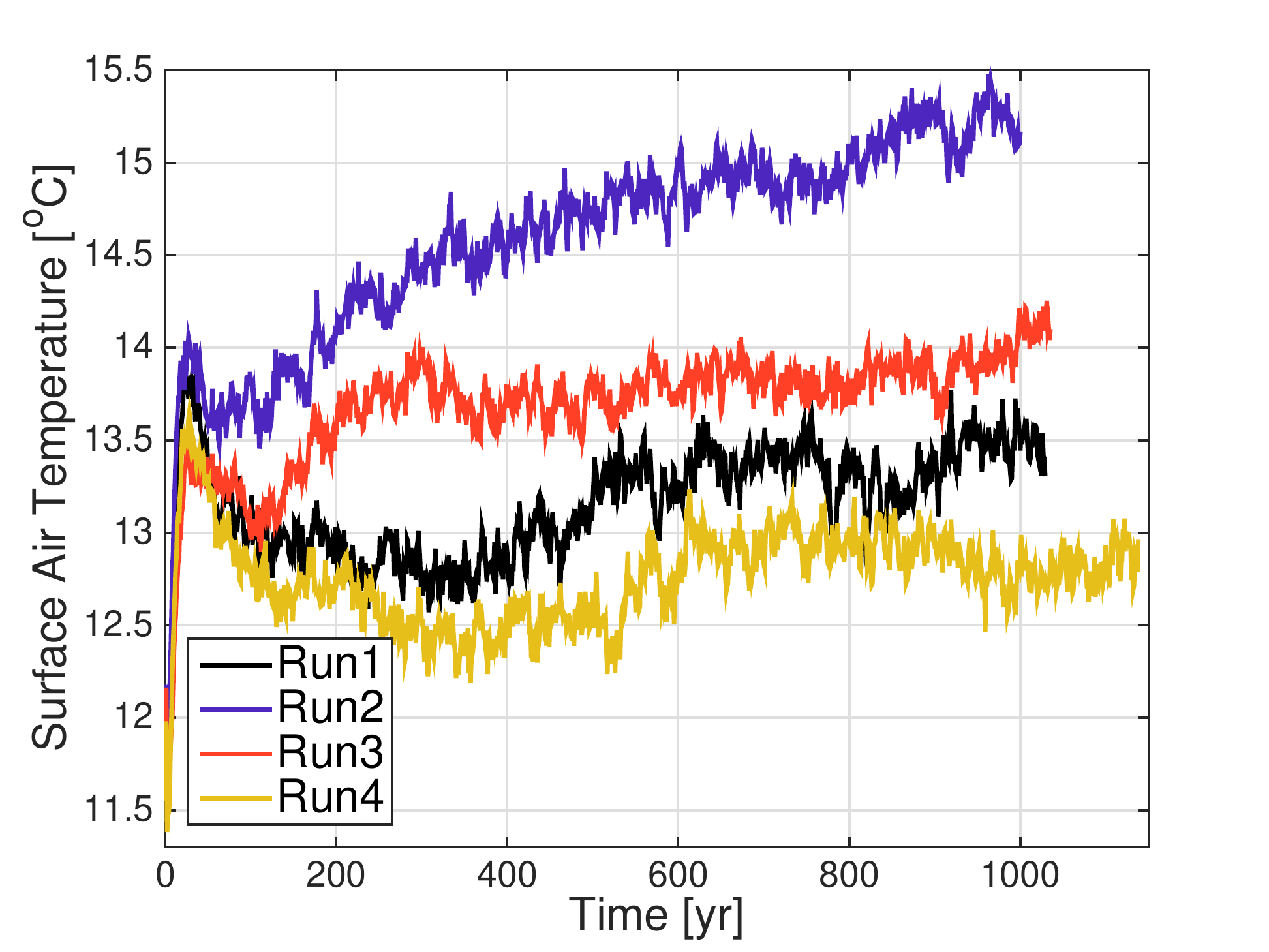}   
\caption{Time evolution of annual-averaged surface air temperature (SAT).}
\label{fig:1}
\end{figure}

\begin{figure}[t]
\includegraphics[width=9cm]{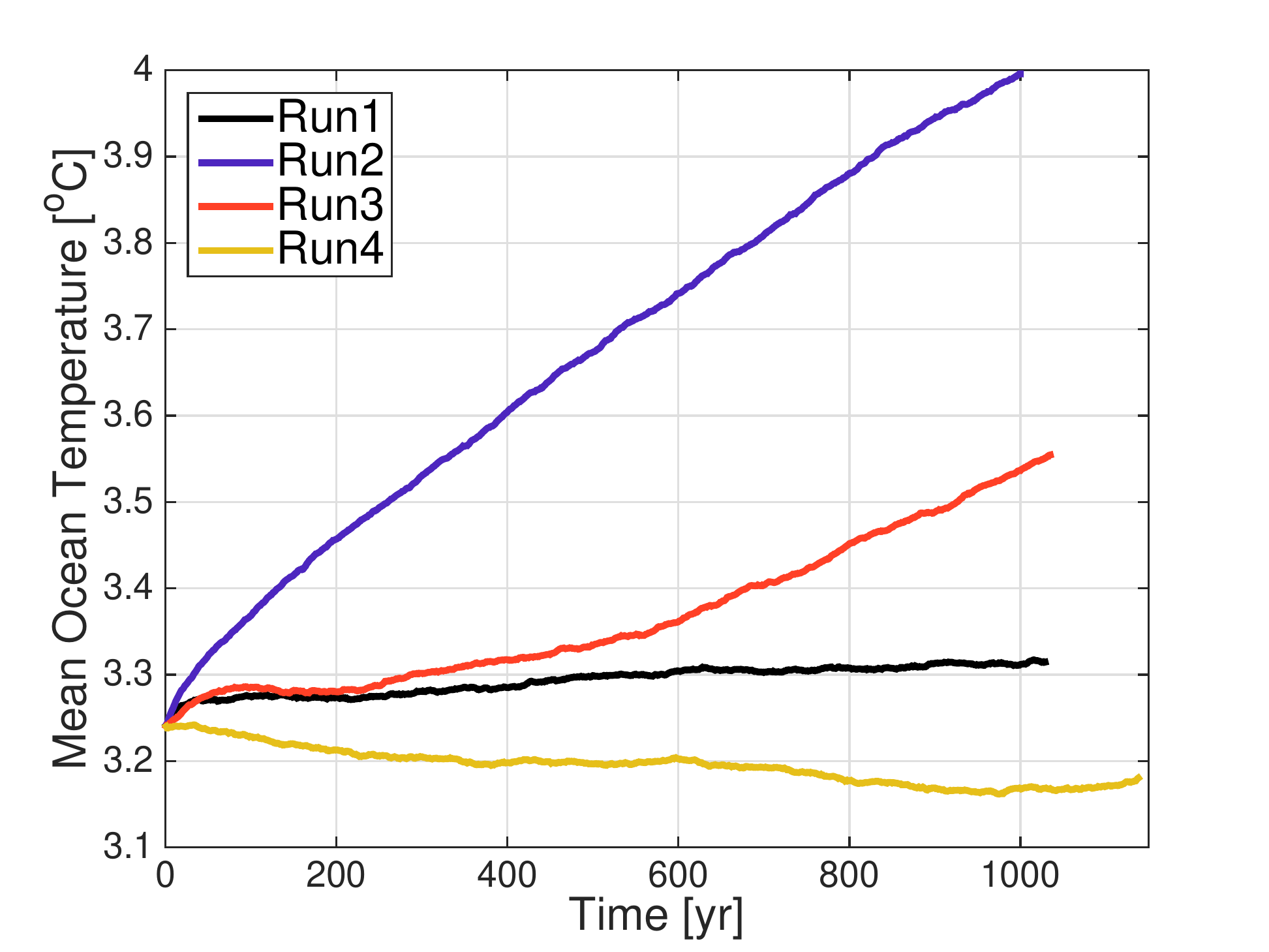} 
\caption{Time evolution of annual-averaged global ocean temperature.}
\label{fig:2}
\end{figure}

\begin{figure*}[t]
\includegraphics[width=8.7cm]{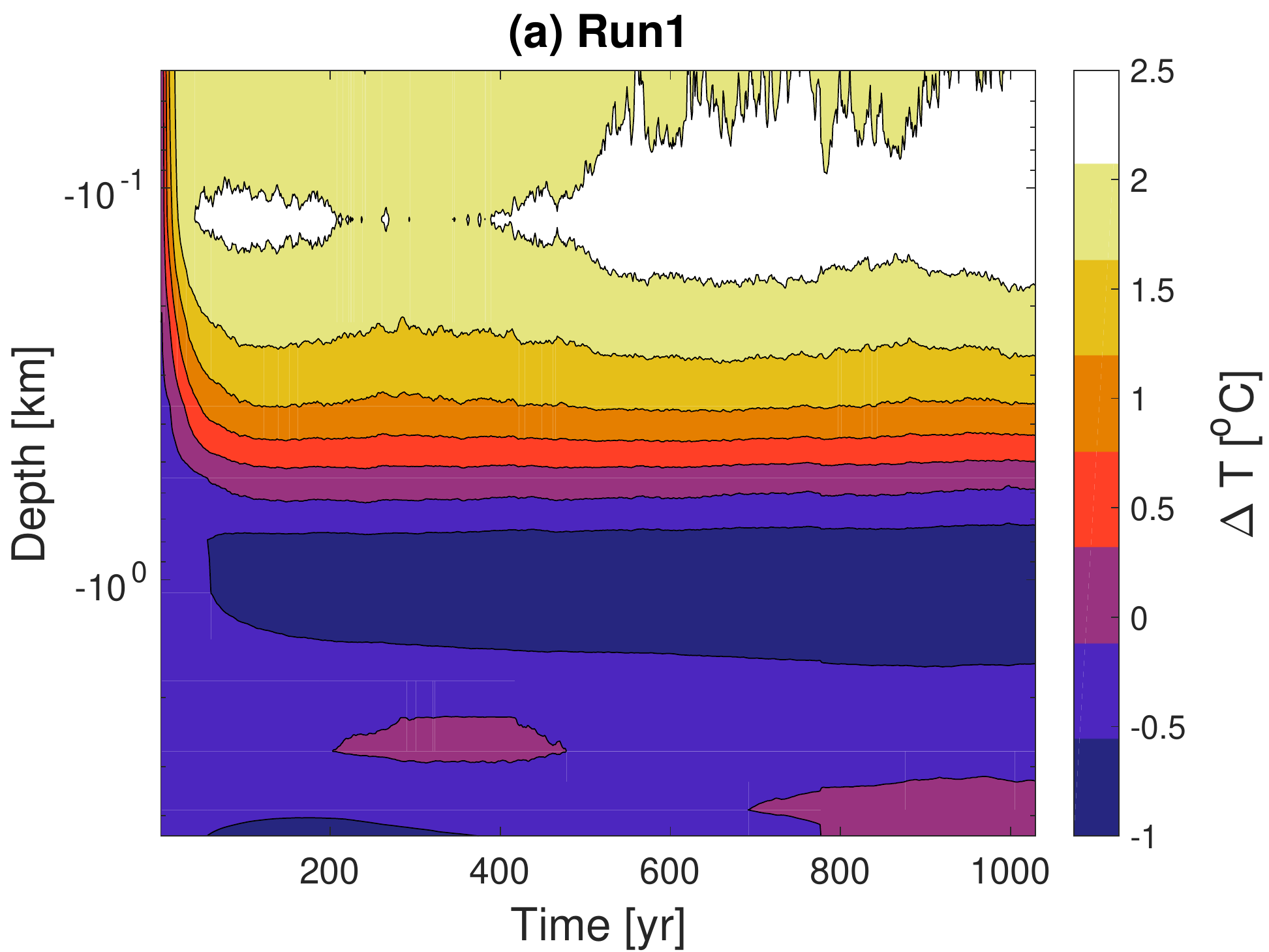}
\includegraphics[width=8.7cm]{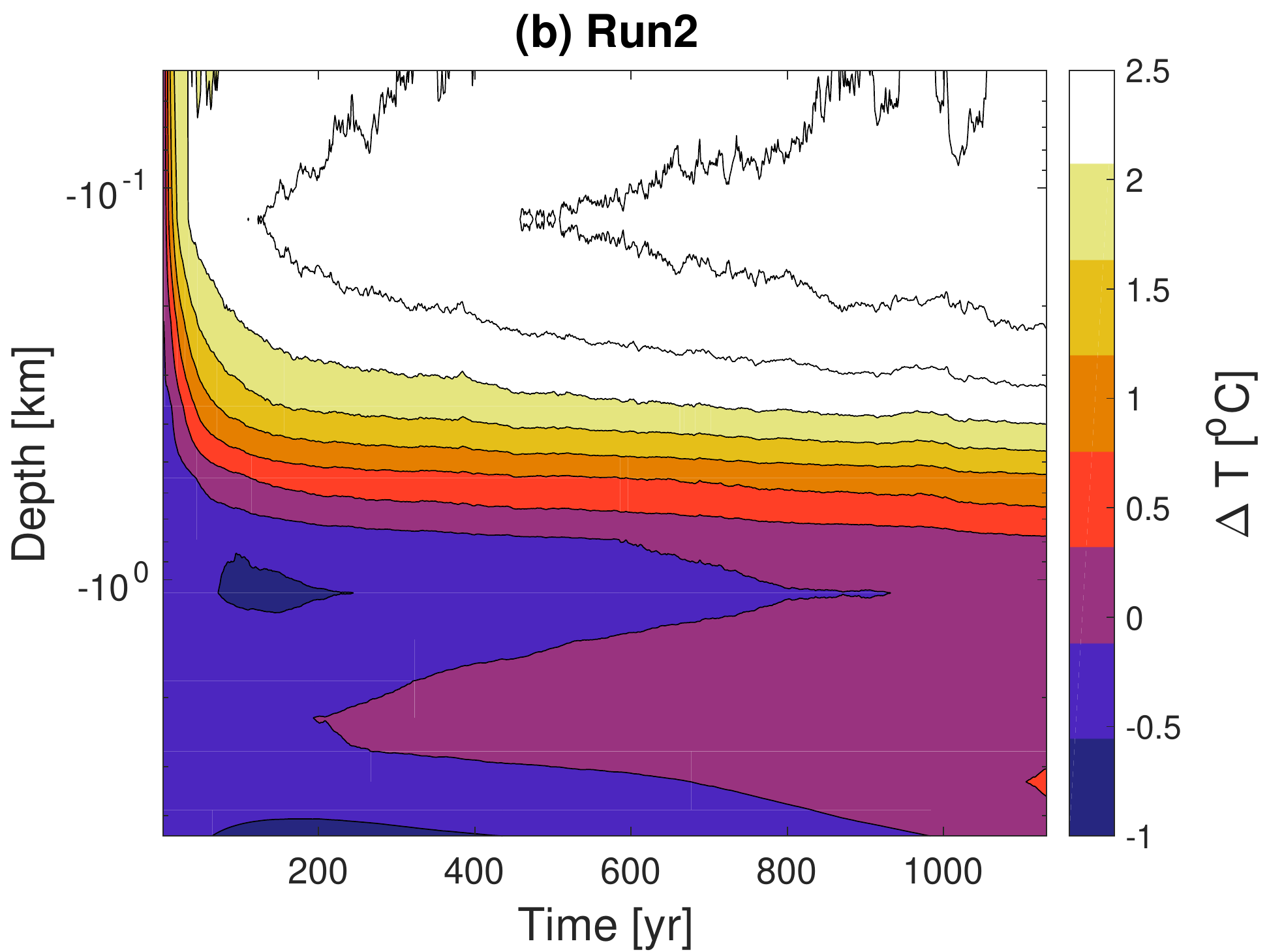}
\includegraphics[width=8.7cm]{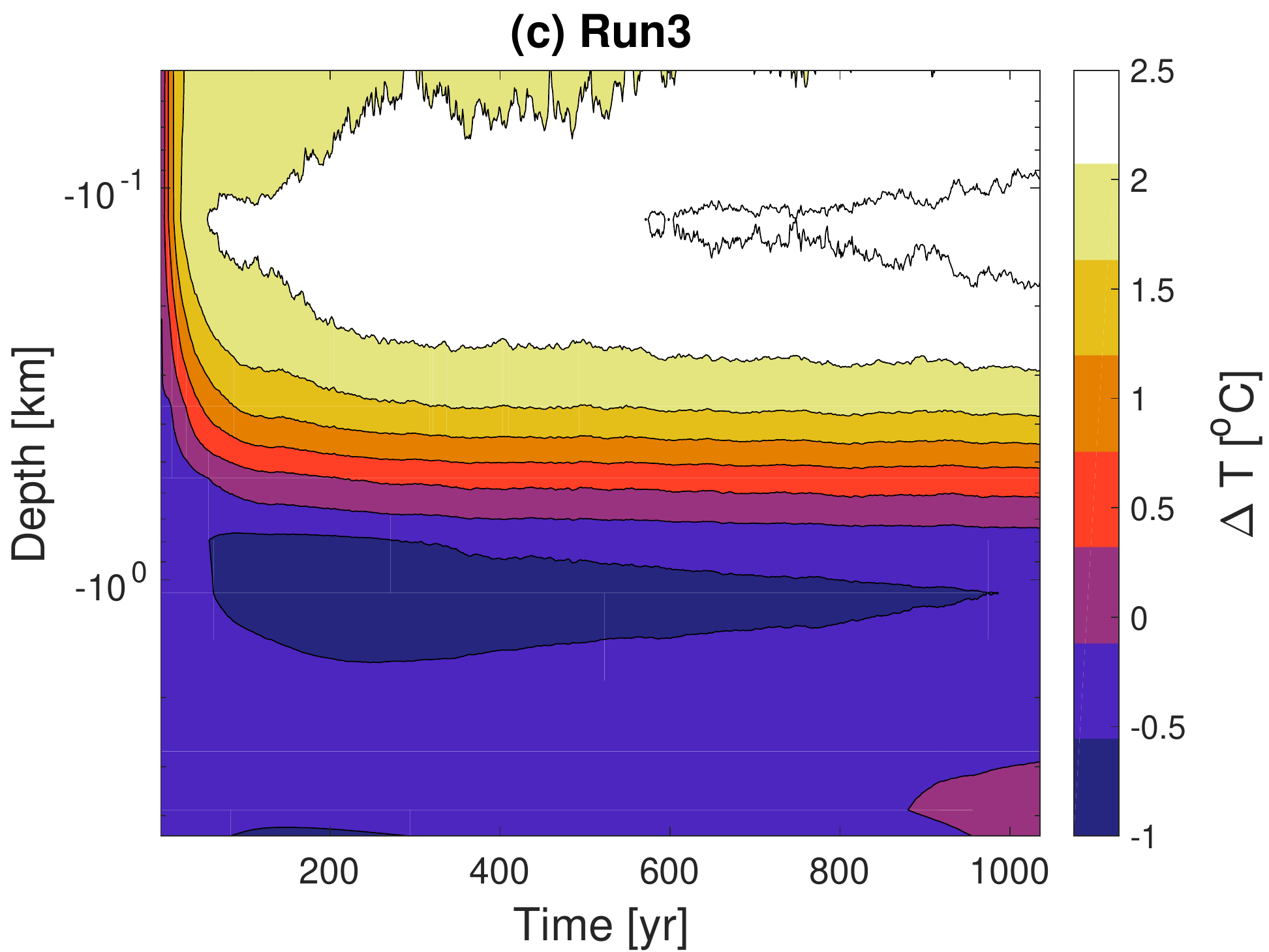}
\includegraphics[width=8.7cm]{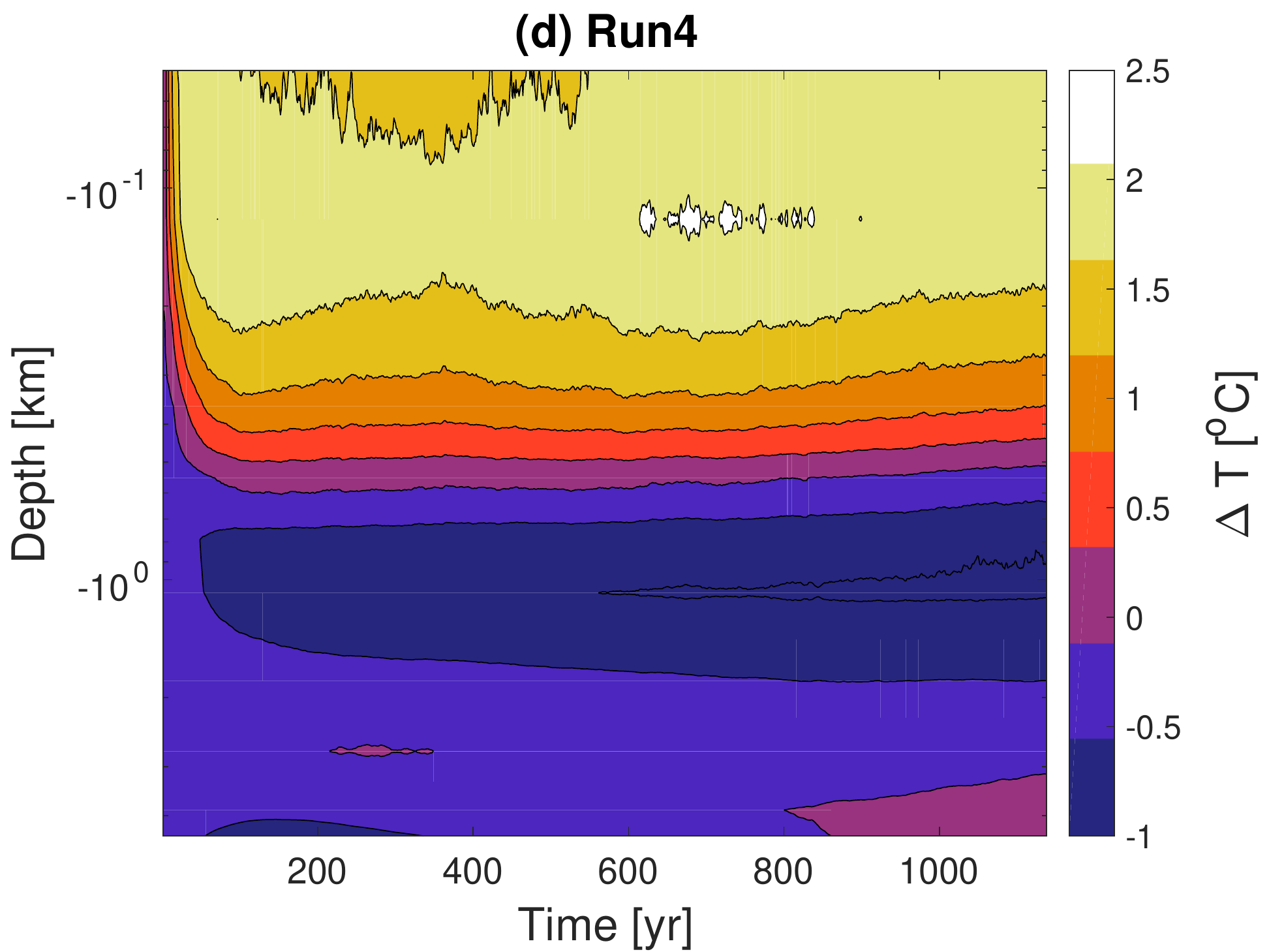}   
\caption{Deviation of the annual-averaged global ocean temperature from the first year as a function of depth.}
\label{fig:3}
\end{figure*}

\begin{figure*}[t]
\includegraphics[width=8.7cm]{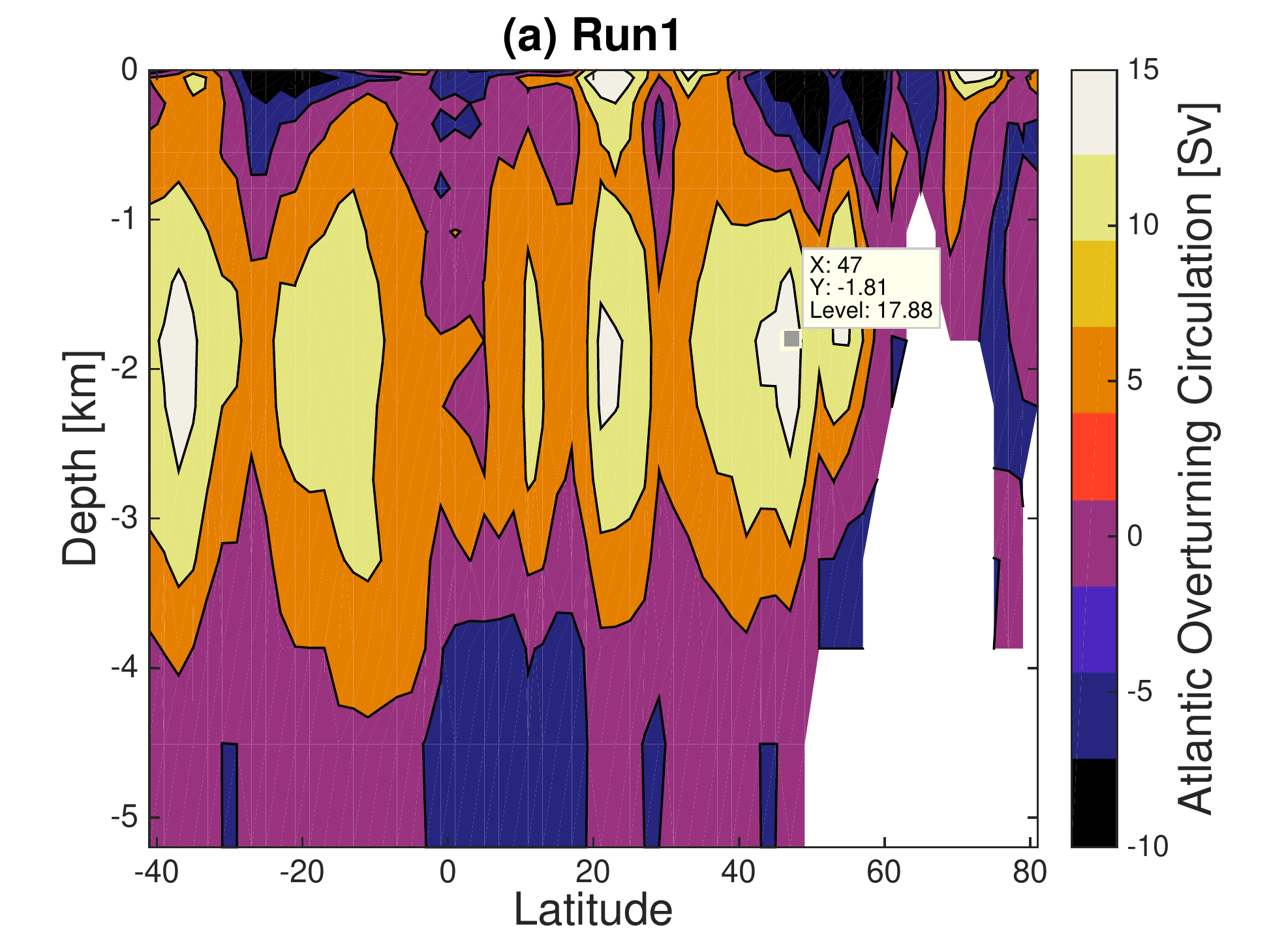}
\includegraphics[width=8.7cm]{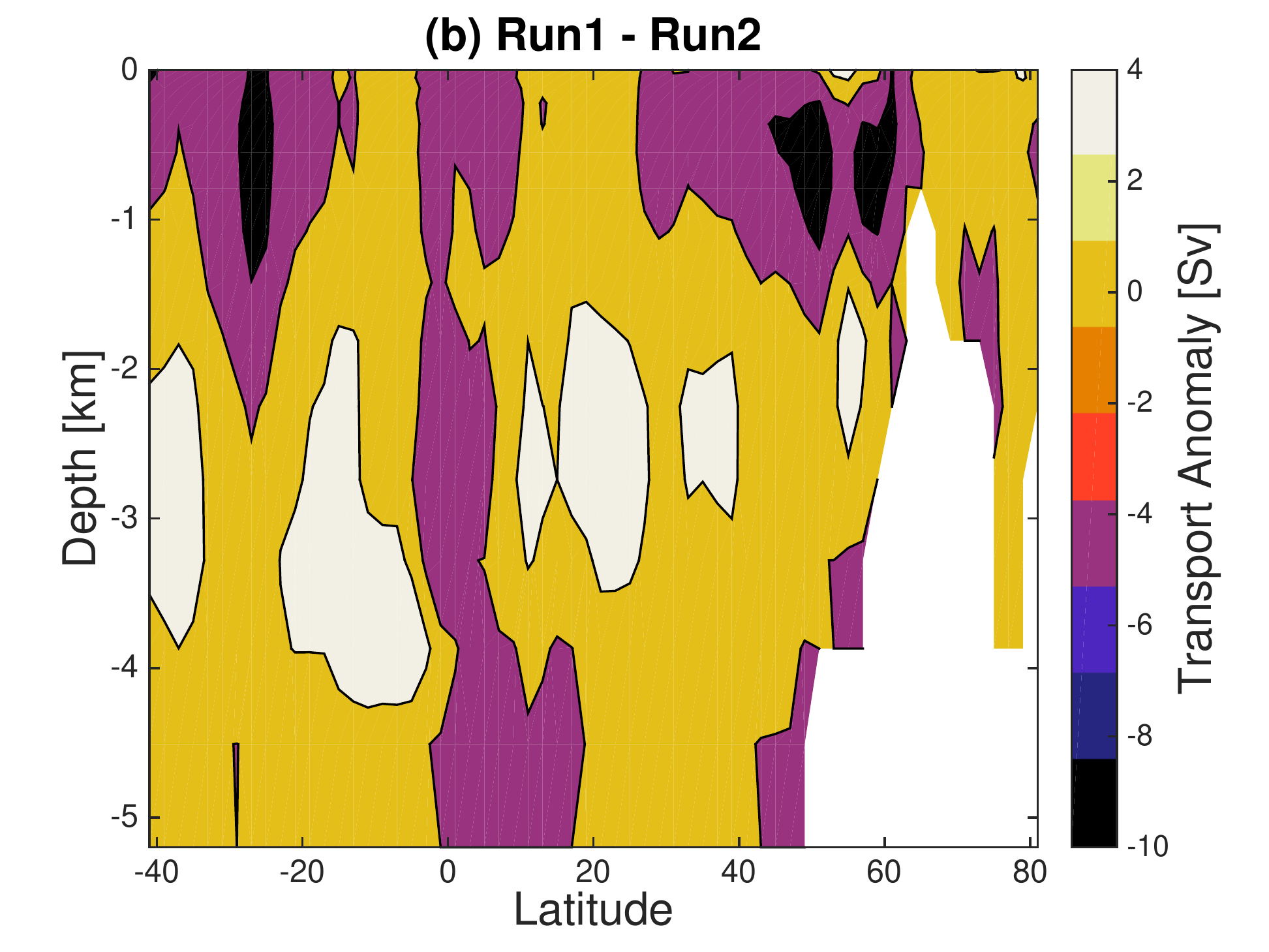}
\includegraphics[width=8.7cm]{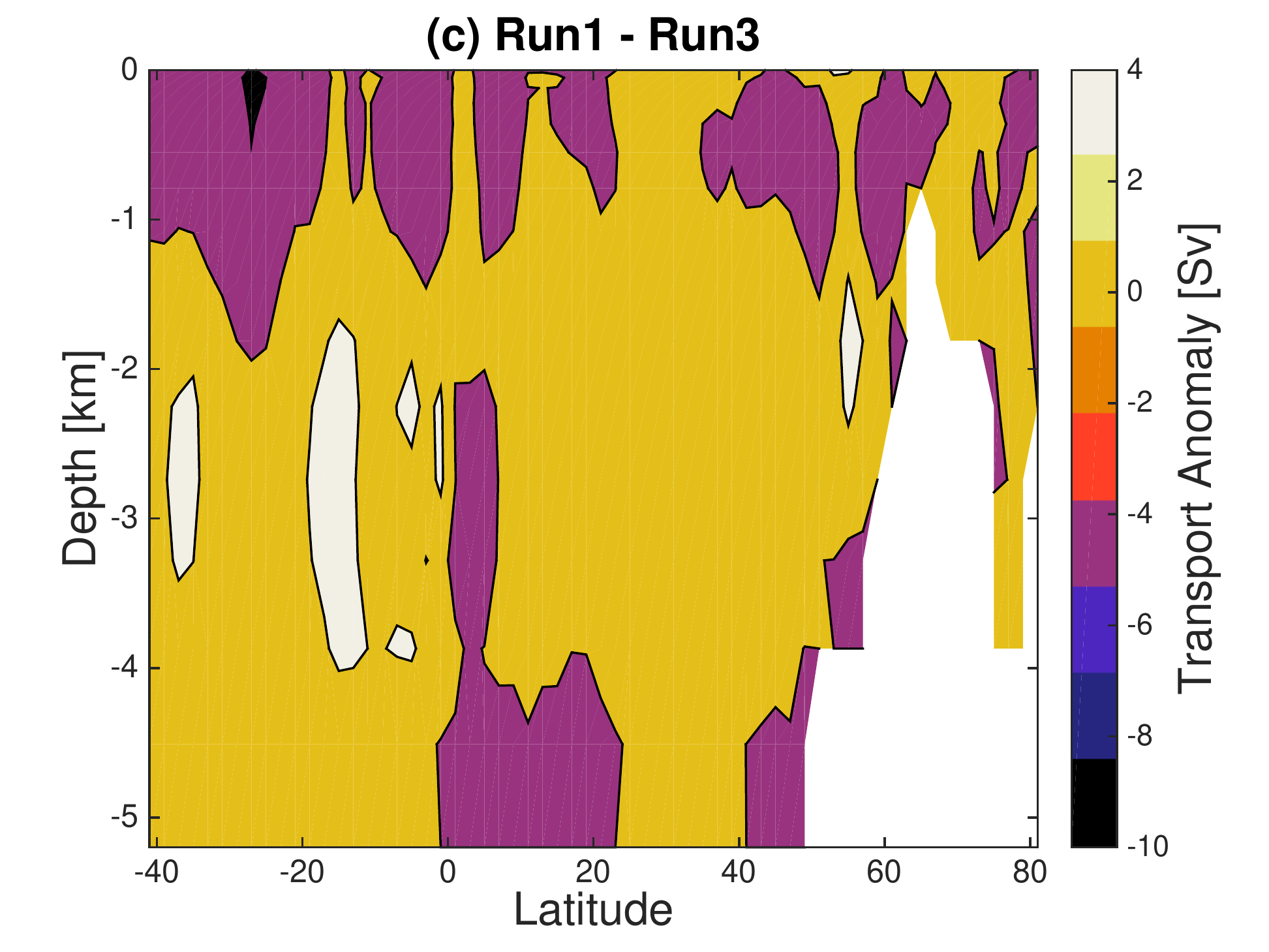} 
\includegraphics[width=8.7cm]{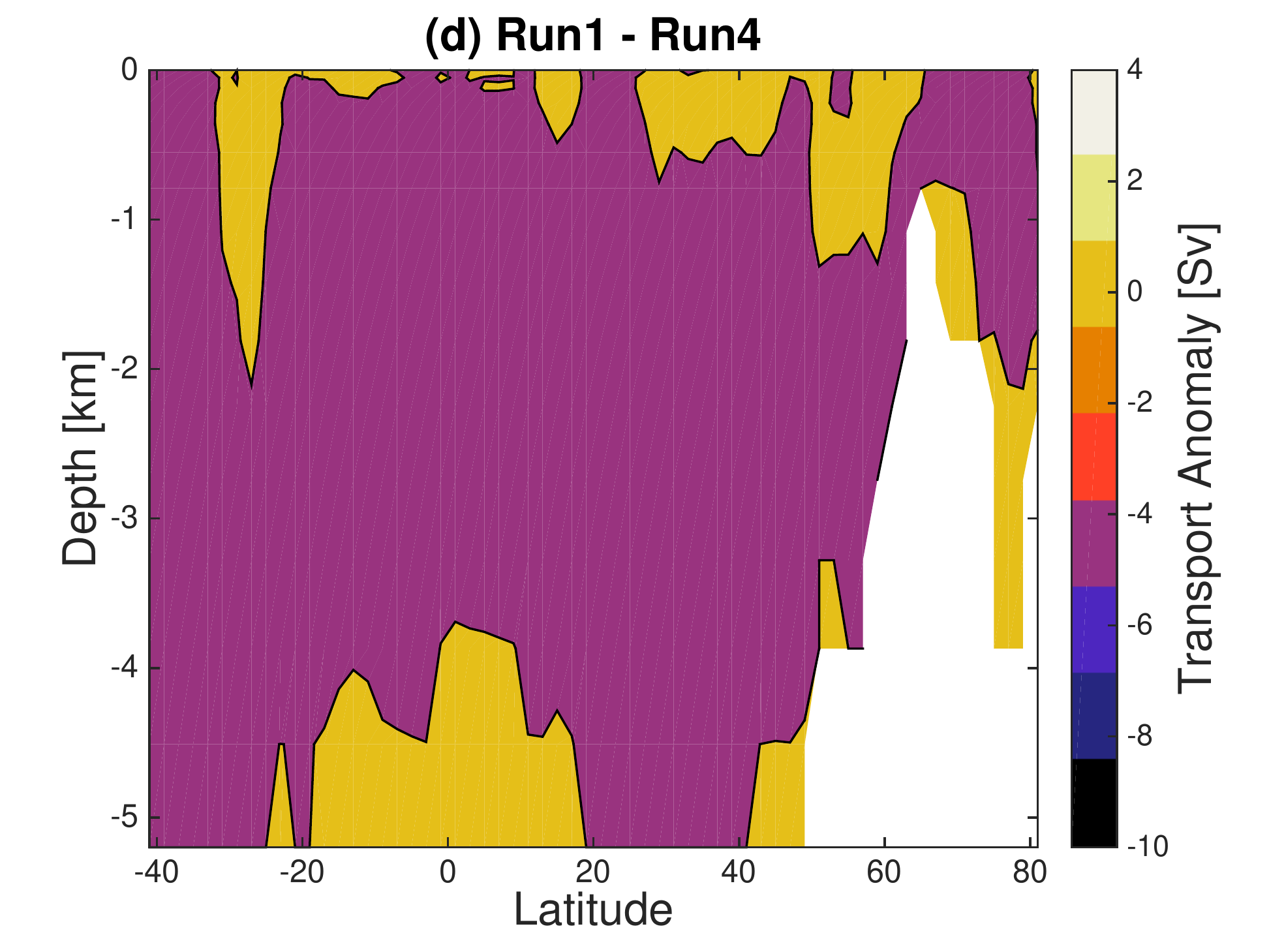}
\caption{Atlantic meridional overturning circulation (AMOC) in Sv $= 10^6$~m$^3$/s as a function of depth and latitude, averaged over the last 100~yr of simulation time. {\it (a)} {\tt Run1}. AMOC anomalies with respect to {\tt Run1} for the simulations {\it (b)} {\tt Run2}, {\it (c)} {\tt Run3} and 
{\it (d)} {\tt Run4}.}
\label{fig:4}
\end{figure*}

The SAT reaches an average value of 13.4$^\circ$C during the last 500 yr of simulation, as can be seen in Fig.~\ref{fig:1}. The global ocean temperature in {\tt Run1} shows a very small drift, with an ocean temperature increase of $(T_{fin}-T_{init})/T_{init}[ ^\circ {\rm{C}}] = $ 2\%  over 1000~yr including the spin-up phase, as shown in Fig.~\ref{fig:2}. The vertical section of the ocean temperature evolution in Fig.~\ref{fig:3}a shows that although during the spin-up phase warming occurs near the surface and cooling at depth, implying vertical redistribution of heat from the deep ocean to the surface, afterwards
the ocean has constant conditions in all the vertical section, including the deep ocean in the last 300~yr. The overturning circulation is well established in the Atlantic ocean with a maximum intensity of 17.9~Sv at latitude 47$^\circ$N, that are typical values in low-resolution runs (see Fig.~\ref{fig:4}a). 

\begin{figure}[t]
\includegraphics[width=8cm]{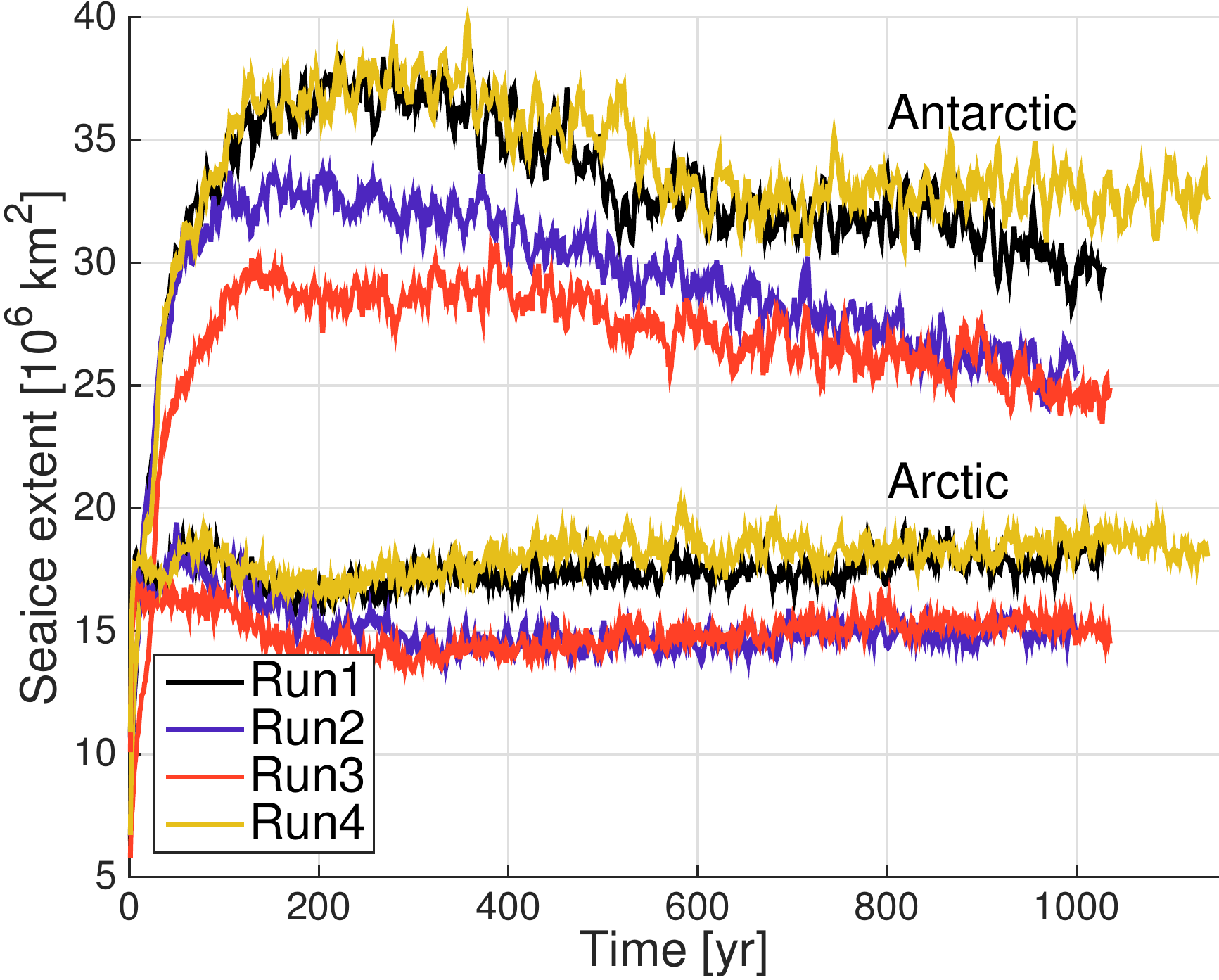}  
\caption{Time evolution of annual-averaged sea ice extent.}
\label{fig:5}
\end{figure}

The Arctic sea ice extent is comparable to annual mean values obtained by other climate models in pre-industrial simulations (ranging from $12.27\cdot 10^6$ to $19.85\cdot 10^6$~km$^2$ in 
\citet{2016CliPa..12..749H}), as can be seen from Fig.~\ref{fig:5}. However, the Antarctic sea ice cover is too extensive as compared to $12\cdot 10^6$~km$^2$ in the observations  and to the annual average of $20.3\cdot 10^6$~km$^2$ obtained in pre-industrial control simulations by CCSM4 \citep{ 2012JCli...25.4817L}, 
probably because the sea ice module in our code does not include dynamical but only thermodynamical effects.
However, the reached values are rather constant during the last 500 yr showing that the simulation is stable during this period of time.  
 
\begin{table*}
\caption{Model parameters used to tune the reference run ({\tt Run1}) and applied to all the other simulations.}
\label{tab:2}
\centering
\begin{tabular}{lc}
\hline\noalign{\smallskip}
Ocean albedo & 0.07 \\
Max sea ice albedo & 0.64 \\
Min sea ice albedo & 0.20 \\
Cold snow albedo &  0.85 \\
Warm snow albedo &  0.70 \\
Old snow albedo & 0.53 \\
Vertical diffusivity & $3\cdot 10^{-5}~{\rm{m}}^2/$s \\
\noalign{\smallskip}\hline
\end{tabular}
\end{table*}

\subsection{{\tt Run2}}
\label{sub:2}

{\tt Run2} is analysed here to illustrate the impact of tuning on the simulation results. The same input parameters used in {\tt Run1} are employed for this simulation except for the value of $RH$ which is set to $RH = 0.85$, a value only 0.7\% larger than the one used in {\tt Run1} (see Table~\ref{tab:1}). The effect of this small increase is that the energy budget at the ocean surface is unbalanced, 0.25~Wm$^{-2}$, giving rise to SAT that reaches 15.2$^\circ$C at the end of the simulation, as shown in Fig.~\ref{fig:1}, and to a huge drift in the global ocean temperature, with an increase of $(T_{fin}-T_{init})/T_{init}[ ^\circ {\rm{C}}] = $ 23\% over 1000 yr (see Fig.~\ref{fig:2}).  In the vertical section (Fig.~\ref{fig:3}b) the ocean temperature increases at all depths (and especially within the first 1000~m depth), while both the intensity of the Atlantic Meridional Overturning Circulation (AMOC) and the sea ice extent decrease with respect to the reference run, as shown in Figs.~\ref{fig:4}b and \ref{fig:5}, respectively.      
We remark that the Antarctic sea ice, continuously decreasing throughout the simulation, is more sensitive to drift than the Arctic sea ice, which stabilises to 
an equilibrium value of $15\cdot 10^6$~km$^2$. 

\begin{figure}[t]
\includegraphics[width=9cm]{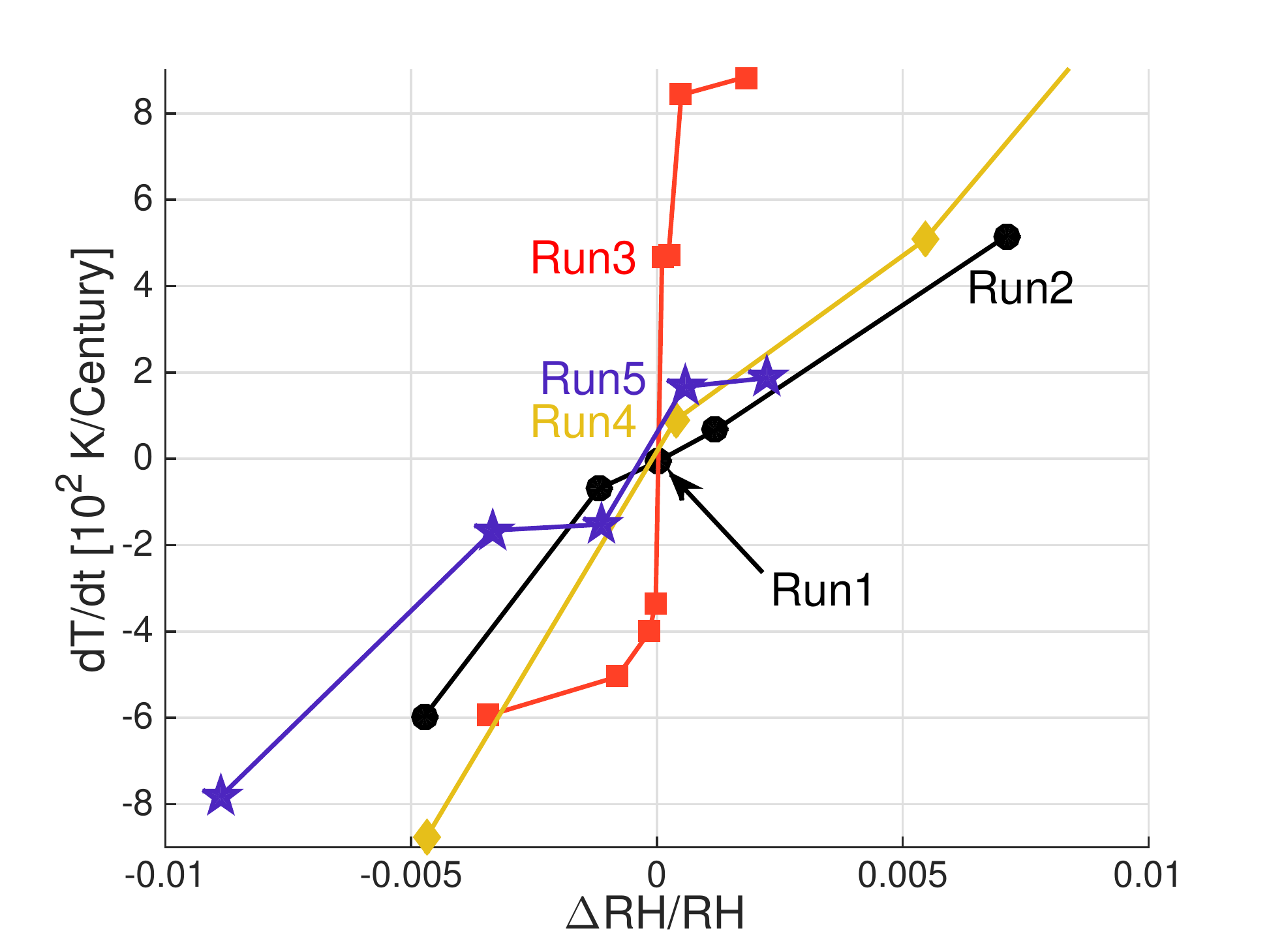}
\caption{Sensitivity to the $RH$ parameter. 
Drift in the mean ocean temperature {\it vs.} the relative error in the $RH$ parameter, $(RH-RH_{c})/RH_{c}$, where $RH_{c}$ corresponds to no-drift, for different modelling set-ups:
8-level SPEEDY cloud-albedo parameterisation as for {\tt Run1} and {\tt Run2} ({\it dots}); 
5-level SPEEDY cloud-albedo parameterisation as for {\tt Run3} ({\it squares}); 
friction heating included as in {\tt Run4} ({\it diamonds}); 
Callovian simulations as {\tt Run5} ({\it stars}). }
\label{fig:6}
\end{figure}

It turns out that there is a strong dependence on the $RH$ parameter that we have summarised in Fig.~\ref{fig:6} by plotting the temperature drift in the ocean as a function of the relative error in this parameter, $(RH-RH_c)/RH_c$, where $RH_c$ corresponds to the no-drift case. Dots in this plot refer to runs with the cloud-albedo parameterisation as in 8-level SPEEDY. We can see that there is a value of $RH$ where the drift is practically equal to zero (this value, $RH = 0.844$, corresponds to the reference run, {\tt Run1}). Small variations around this value gives rise to unbalanced runs, like {\tt Run2}.    

Note that drift in the global ocean temperature occurs in many of the state-of-the-art climate model simulations conducted under CMIP5~\citep{2016JCli...29.1639H} where $dT/dt$ is typically higher than the value reached in {\tt Run1}.  Given that models drift away from the observed state chosen at initialisation if tuning is not sufficiently accurate, short spin-up results in greater fidelity to initial conditions with respect to long spin-up. On the other hand, long spin-up is an important factor 
in drift reduction~\citep{SenGupta2012,SenGupta2013}. 
Our analysis shows that choosing a value of $RH$ that minimises the 
drift, such in {\tt Run1}, increases model fidelity to initial conditions and at the same time guarantees stability over simulation times of the order of thousand years. 

\begin{table*}
\caption{Energy imbalances and drift in the global ocean temperature.}
\label{tab:3}
\centering
\begin{tabular}{cccc}
\hline  
Simulation  & TOA imbalance & Surface imbalance  & $dT/dt$ \\
 &    [Wm$^{-2}$] &  [Wm$^{-2}$] &   [K/Century]\\
\hline 
{\tt Run1} &   2.65 & 0.00 & 0.001 \\
{\tt Run2} &    2.81 & 0.25 & 0.052 \\
{\tt Run3} &   2.74 & 0.23 & 0.047 \\
{\tt Run4} & -0.55 & 0.04 & 0.009 \\
{\tt Run5} &   -0.36 & 0.09 &  0.016\\
\hline
\end{tabular}
\end{table*}

\subsection{{\tt Run3}}
\label{sub:3}

We have tested in {\tt Run3} a different way of representing the bulk cloud albedo, to be compared to the one used in {\tt Run1}. As in the atmospheric module 5-level SPEEDY~\citep{Molteni2003}, the bulk cloud albedo is held constant in {\tt Run3}. Since this parameterisation gives rise to a solar radiation that is too strong at high latitudes in coupled models, developers of the SPEEDY code remarked that using a bulk cloud albedo that increases with latitude improved the simulation results~\citep{2016ClDy...46.2337K}. We thus expect that the parameterisation used in {\tt Run3} is less accurate than the one used in the reference run ({\tt Run1}).  

\begin{figure}[t]
\includegraphics[width=9cm]{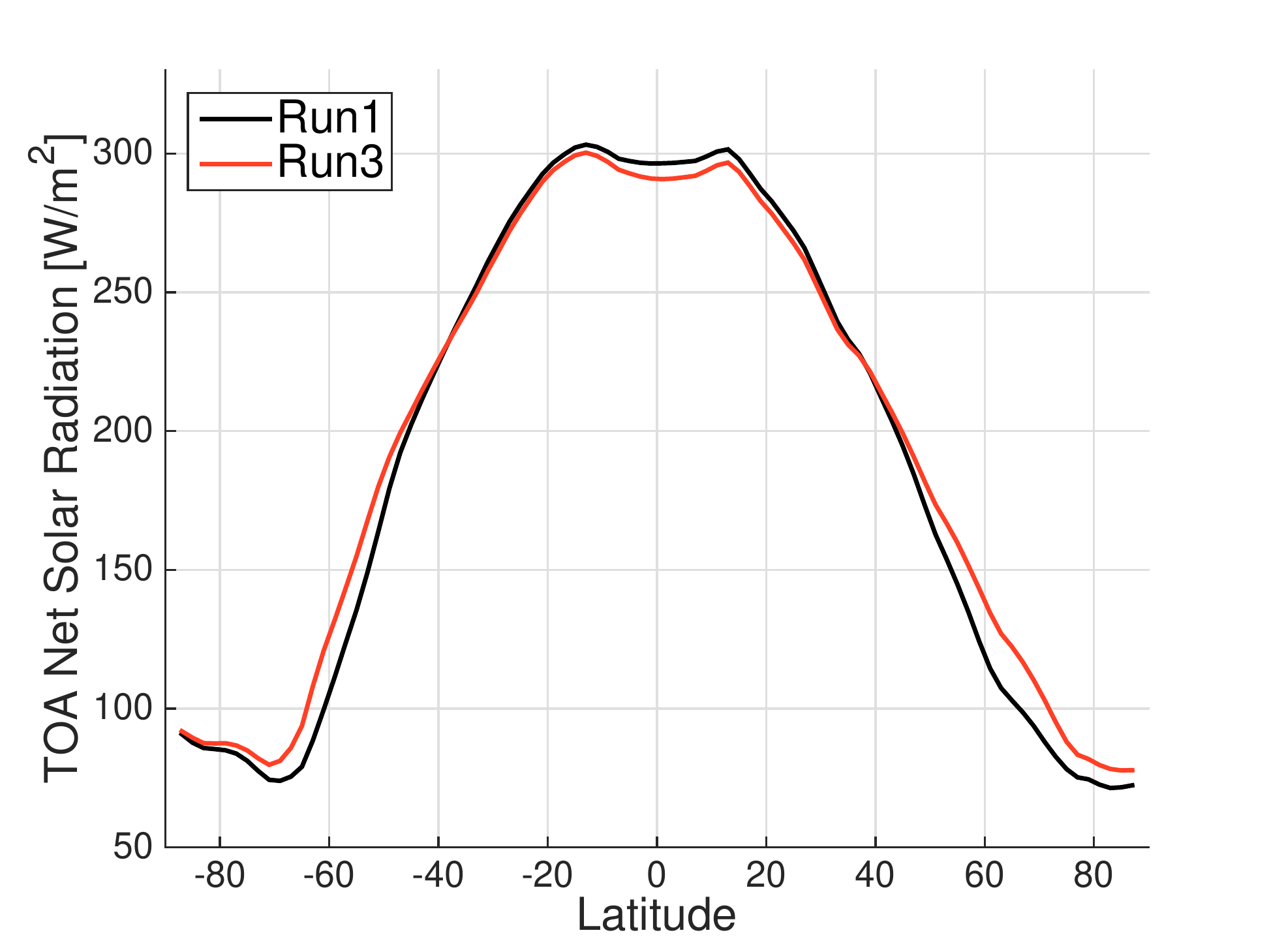}  
\caption{Zonal mean of TOA net shortwave radiation in simulations {\tt Run1} and {\tt Run3}. }
\label{fig:7}
\end{figure}

We have checked that the parameterisation used in {\tt Run1} produces less net solar radiation at high latitudes (see Fig.~\ref{fig:7}) with respect to {\tt Run3}, as expected from the different description of bulk cloud albedo employed in 8-level and 5-level SPEEDY.
We find that while SAT is rather constant during the first 800 yr in {\tt Run3}, it tends to increase in the last century (see Fig.~\ref{fig:1}). The annual-averaged global ocean temperature has a positive trend all along the simulation (see Fig.~\ref{fig:2}). This increase in temperature can also be observed in Fig.~\ref{fig:3}c where the ocean temperature is seen to raise, especially in the upper 500~m. This change in the ocean and surface air temperature can be related to a decrease of sea ice extent in both north and south polar regions with respect to the reference run, as can be seen in Fig.~\ref{fig:5}. 
The AMOC is only slightly affected, as can be seen in 
Fig.~\ref{fig:4}c, while the total heat transport is smaller than in the case of {\tt Run1}, as can be 
seen in Figs.~\ref{fig:8}a. 

The monotonic increase in global ocean temperature de\-monstrates that {\tt Run3} is not well stabilised. It is important to note that if the tuning parameter is slightly reduced, the result is a monotonic decrease in the global ocean temperature, showing that it is very difficult to obtain  a stable control run for this series of simulations. 
In order to better understand this point, we have investigated the behaviour of the temperature drift as a function of the relative error in $RH$ parameter (Fig.~\ref{fig:6}, squares) and interestingly we found that the dependence on this relative error is different from the case of {\tt Run1}-{\tt Run2} series (Fig.~\ref{fig:6}, dots). In the present case, the temperature drift $dT/dt$ is strongly sensitive to the $RH$ parameter: a huge change in $dT/dt$ (of the same order as that occurring between {\tt Run1} and {\tt Run2}) is obtained for very small variations of $RH$ (of the order of 
0.01\%, to be compared with 0.7\% in the case of {\tt Run1} and {\tt Run2}). This means that the parameterisation considered in {\tt Run3} and all the runs in the same series represented by squares in Fig.~\ref{fig:6} is much more sensitive to the tuning parameter $RH$ than the parameterisation used in {\tt Run1} series (dots in Fig.~\ref{fig:6}). This sensitivity can be used as a criterium to establish for the goodness of a given parameterisation since the more the 
sensitivity to the tuning parameter,  the smaller the range where the tuning parameter can vary to obtain minimal drift and well-stabilised control runs.  
This criterium, applied to different tuning parameters and/or parameterisations, can be generalised to other coupled climate models.   

\subsection{{\tt Run4}}
\label{sub:4}

The origin of the imbalance at TOA has received a great deal of attention from the scientific community~\citep{2011ClDy...36.1189P,2011RvGeo..49.1001L,2016JCli...29.1639H}. 
The imbalance is ascribed to physical processes that have been neglected or approximated in climate models. In our set-up, the main source/sink of heat 
are {\it (i)} the neglect of the heating due to kinetic energy dissipation by internal stress and viscous processes~\citep{2014ClDy...43..981L} ; {\it (ii)} 
the approximation of considering fixed soil moisture; {\it (iii)} the neglect of the sea ice dynamics; {\it (iv)} 
the approximation used in the thermodynamical module to limit the ice thickness to a maximal value of 10~m. Other sources of imbalance have numerical origin and can be related to the time-stepping method, to the advection schemes or to hyperdiffusion operators introduced in the dynamical core in order 
to smooth and avoid divergences~\citep{2014ClDy...43..981L}.      

The heating due to kinetic energy dissipation is in general dominant with respect to the other effects and can reach values of the order of 3~Wm$^{-2}$ in other coupled atmosphere-ocean general circulation models, such as Had\-CM3 and its coarse-resolution version FAMOUS~\citep{2011ClDy...36.1189P}. 
Mo\-reover, biases ranging from $-0.3$ to 4.8~Wm$^{-2}$ are found in the net global balance of pre-industrial simulations included in the IPCC4AR~\citep{2011RvGeo..49.1001L}. 
In order to better understand this point, we have performed simulations where the friction heating is returned back to the atmospheric component by activating the corresponding switch in MITgcm.  
We obtain a simulation, {\tt Run4}, where the TOA imbalance is much reduced with respect to the reference run, as can be seen in Table~\ref{tab:3}. The tuning 
procedure gives rise to a very well equilibrated simulation, as can be inferred from the time evolution of SAT, that reaches a rather constant value of 12.8$^\circ$C  (Fig.~\ref{fig:1}), and from the global ocean temperature, that decreases of only $(T_{fin}-T_{init})/T_{init}[ ^\circ {\rm{C}}] = -2$\% in 1100~yr (Fig.~\ref{fig:2}). 
From the vertical section of the ocean temperature shown in Fig.~\ref{fig:3}d, we observe that there is less  warming of the upper ocean 
than in {\tt Run1}, the AMOC increases almost everywhere by 4~Sv (Fig.~ \ref{fig:4}d) while the sea ice extent becomes very well stabilised in the last 500~yr of the simulation (Fig~\ref{fig:5}).

\begin{figure}[t]
\includegraphics[width=8cm]{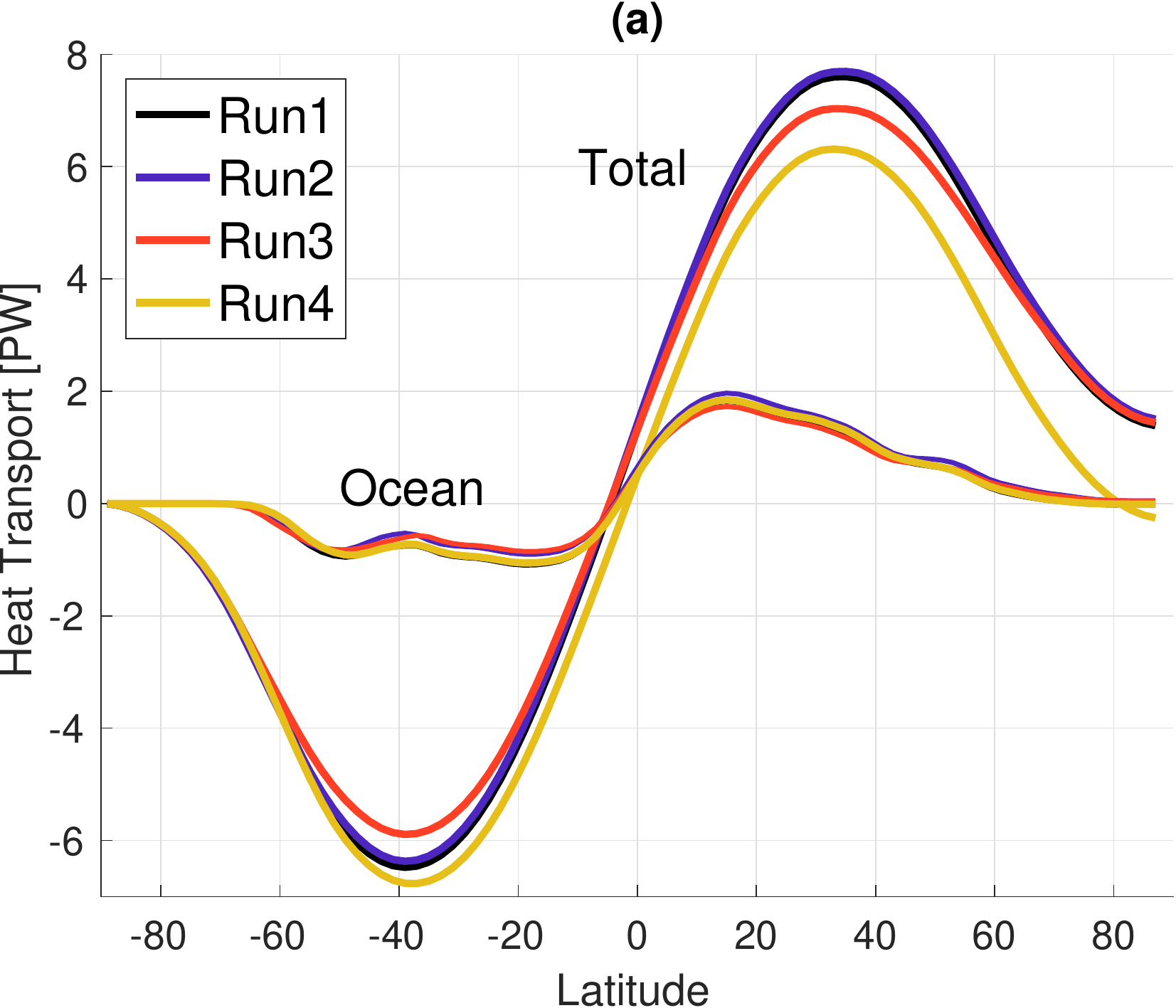}  
\includegraphics[width=8cm]{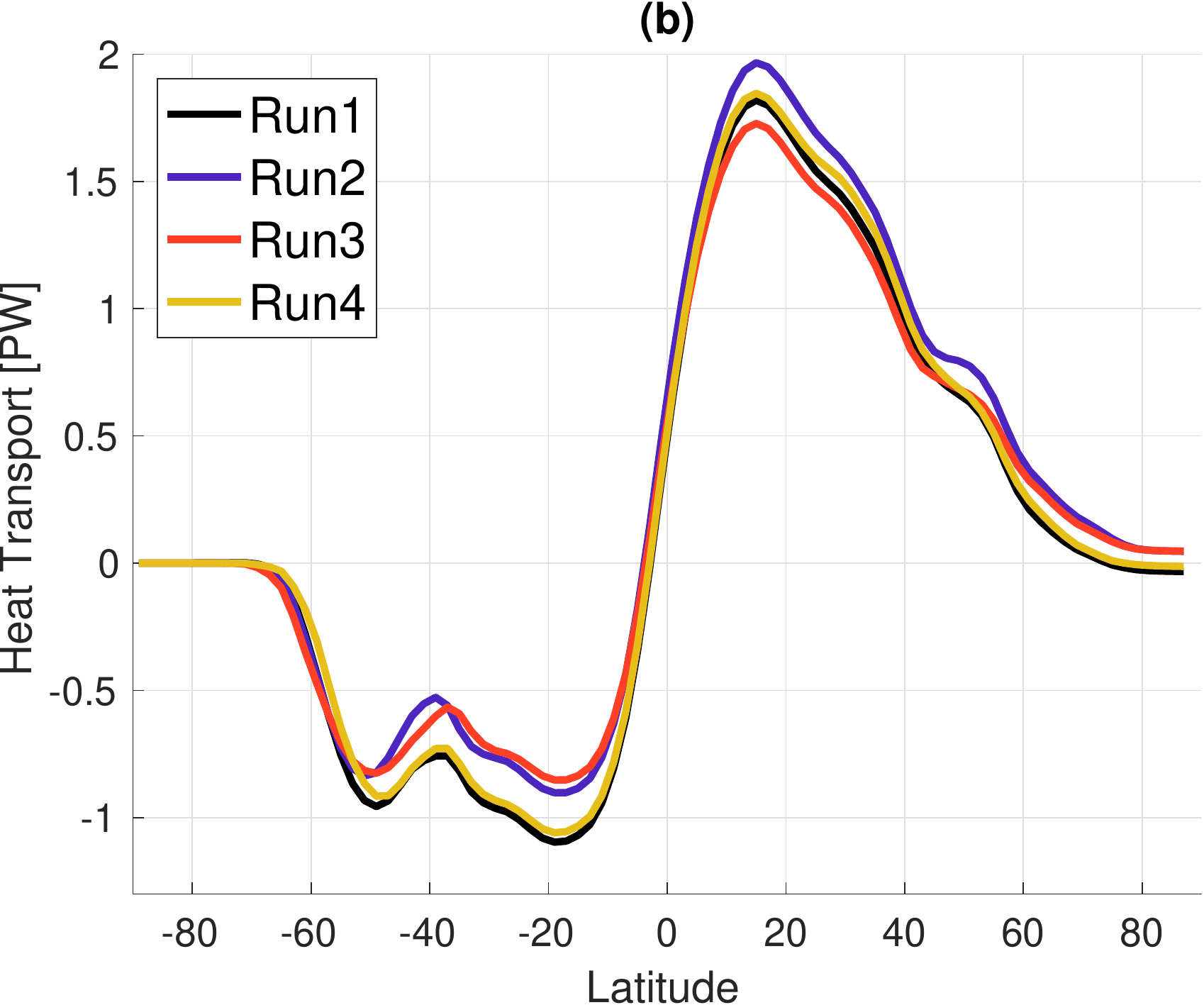}  
\caption{{\it (a)} Total heat transport in PW ($10^{15}$~W) and the oceanic contribution to this total transport. A positive value on the vertical axis corresponds to a northward transport. {\it (b)} Zoom of the oceanic contribution.}
\label{fig:8}
\end{figure}

The range where the tuning parameter gives reasonable well equilibrated runs is comparable to {\tt Run1} series (compare the slope of the curves with dots and diamonds in Fig.~\ref{fig:6}), thus we can conclude that including friction heating back into the atmospheric column has a global positive effect on the quality of the simulations, since the TOA imbalance is reduced with respect to {\tt Run1}, and this is a result that can be generalised to other coupled climate models. The positive effect of accounting for the friction heating clearly appears in the total heat transport, 
that is much closer to observed values of the order of 6~PW at 30$^\circ$N~\citep{2008JCli...21.2313F} in {\tt Run4} than in all the other considered simulations, as shown in Fig.~\ref{fig:8}. 
The improvement is particularly effective 
in the atmospheric contribution to the total heat transport especially within the northern hemisphere, while it is negligible in the oceanic contribution (compare {\tt Run1} and {\tt Run4} in Figs.~\ref{fig:8}a and b). 

\subsection{{\tt Run5}}
\label{sub:5}

\begin{figure*}[t]
\includegraphics[width=8.7cm]{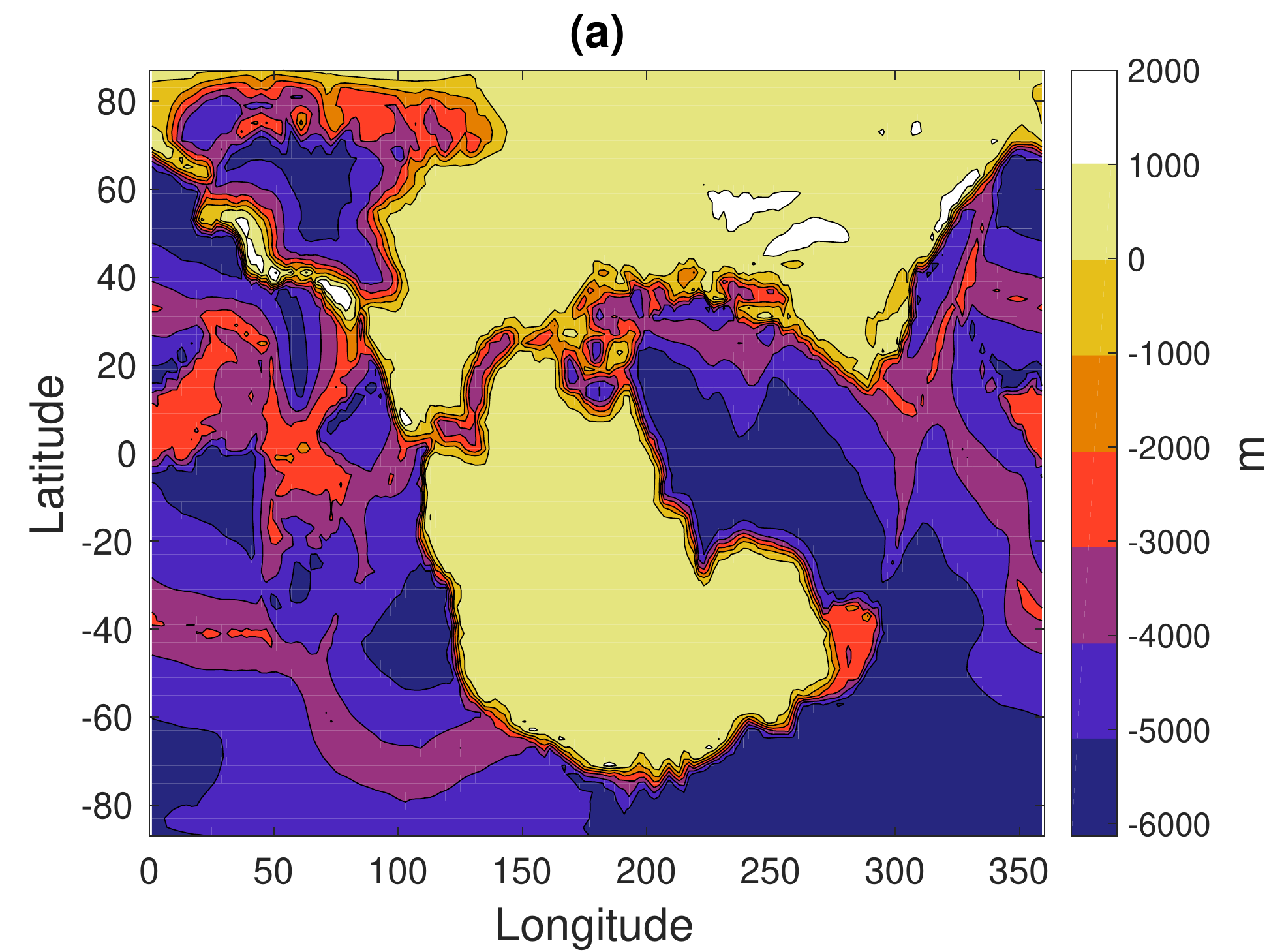}
\includegraphics[width=8.7cm]{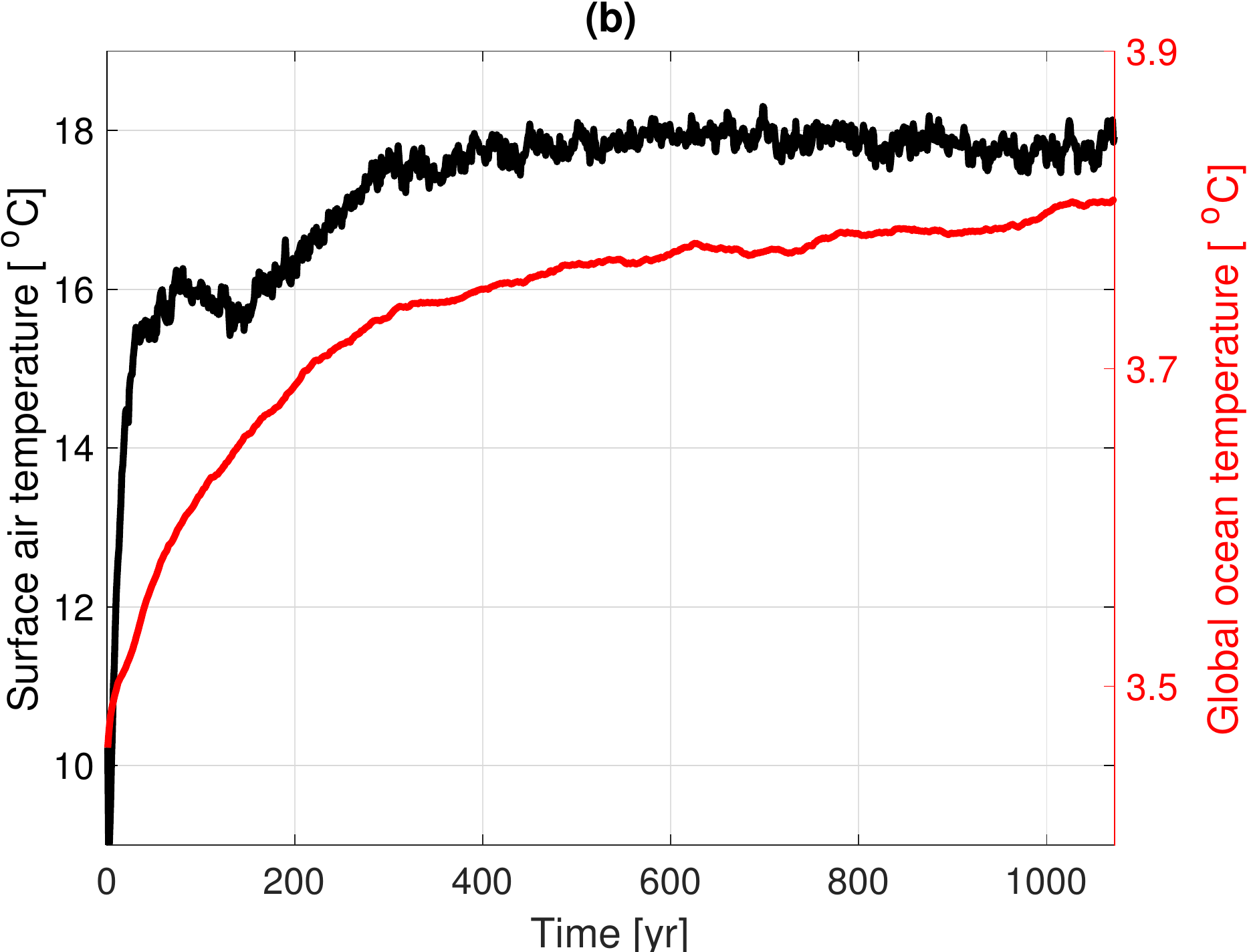}
\includegraphics[width=8.7cm]{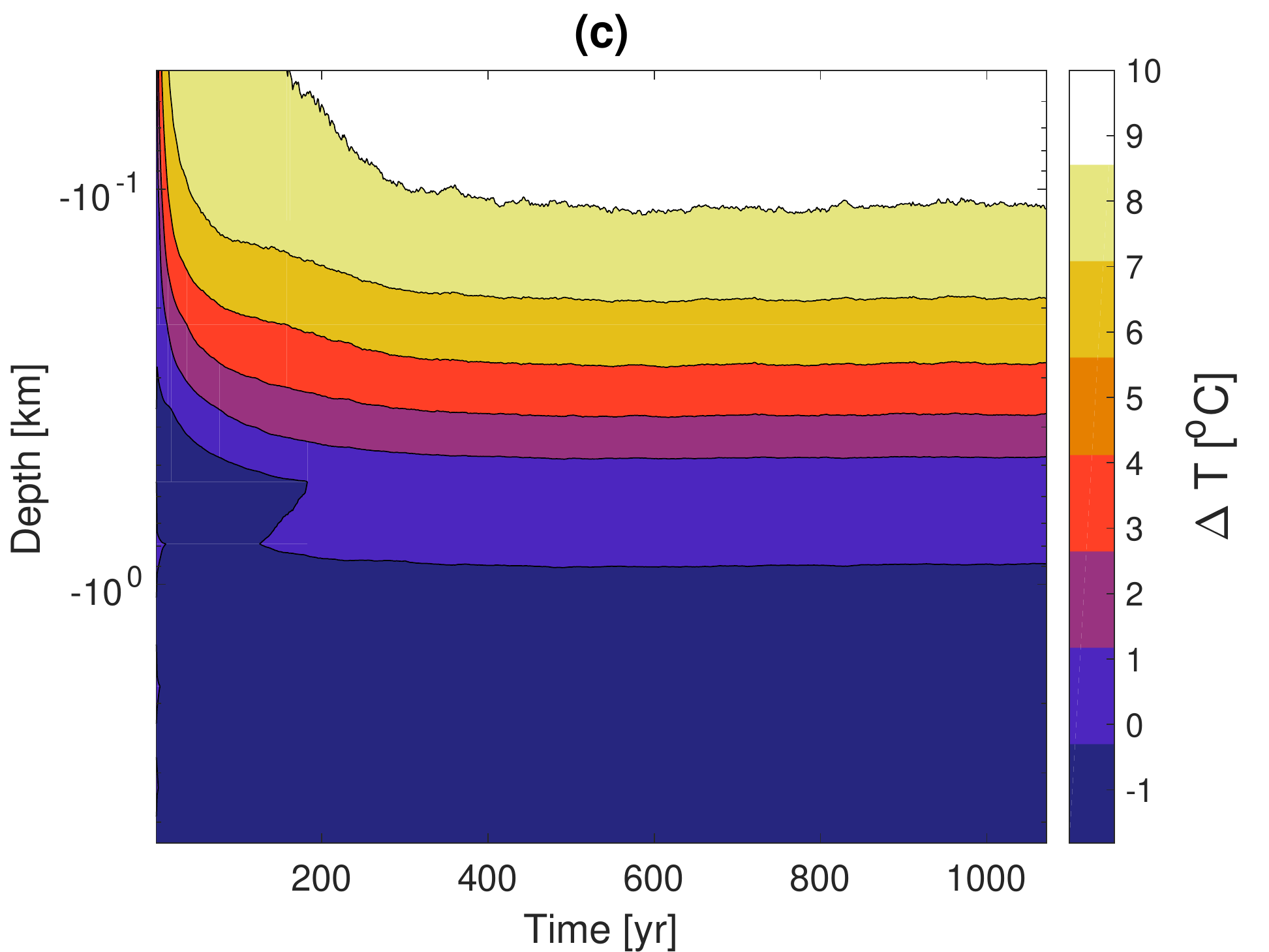}
\includegraphics[width=8.7cm]{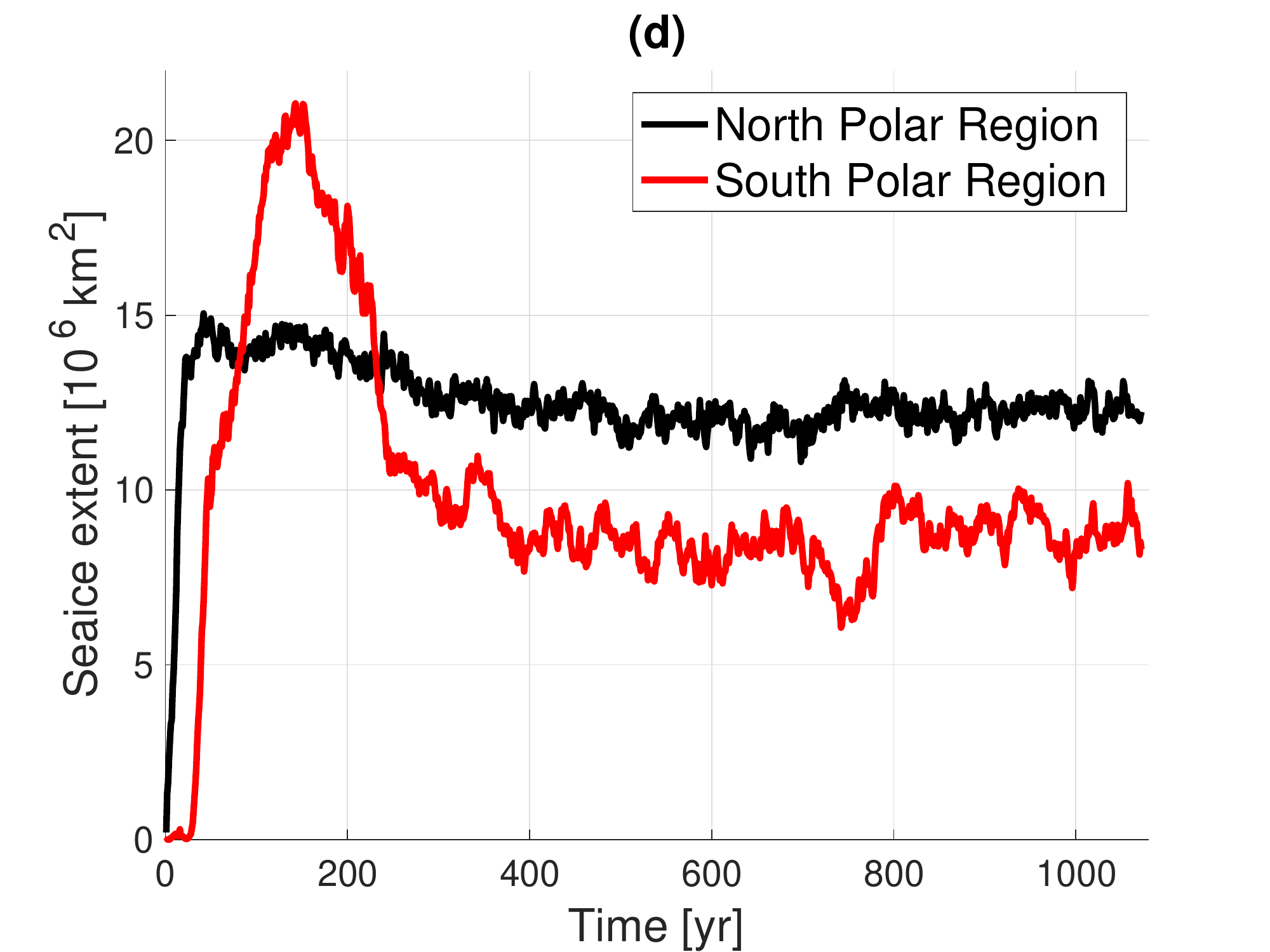}
\caption{Callovian simulation {\tt Run5}. {\it (a)} paleogeography. {\it (b)} Time evolution of annual-averaged surface air temperature 
and global ocean temperature. {\it (c)} Deviation of the annual-averaged global ocean temperature from the first year as a function of depth. {\it (d)} Time evolution of annual-averaged sea ice extent.}
\label{fig:9}
\end{figure*}

We now apply the previous method to the case of a different paleogeographic configuration. Since we have seen that including friction heating back into the atmospheric component had positive effects on the simulation results, we employ the same procedure here. We also use the cloud albedo parameterisation of {\tt Run1} that gives improvements with respect to the one used in  {\tt Run3}.  The paleogeographic configuration that we consider here corresponds to the Callovian-Oxfordian transition  (165~Ma) (see Fig.~\ref{fig:9}a or, for more details, Fig.~1 in~\citet{BrunettiVerard2015}), where Pangea breakup gave rise to the formation of the Central Atlantic and the Proto-Caribbean basin, and provided a connection between the Neo-Tethys and Panthalassa.  
We have already used this configuration in ocean-only simulations presented in~\citet{BrunettiVerard2015}. Apart from the interest of studying this 
period using a coupled climate model and comparisons with geological data, that we will pursue in forthcoming publications, 
the technical challenge is now to obtain a stable simulation 
by tuning the $RH$ parameter so that the TOA energy imbalance is minimal and the surface energy imbalance is nearly zero.      
For the present test we do not put much effort in optimising the initial conditions, for which we took the zonal average of the potential temperature over the present-day Pacific ocean, a constant 
salinity distribution at 39~psu, a vegetation cover that is homogeneous in latitude from~\citet{2008ProcGeol..119..5P}, and a runoff map based on the topography used in~\citet{BrunettiVerard2015}. 
The initial conditions are anyway affected by many uncertainties, the range of published estimates of Jurassic atmospheric CO$_2$ content varying between present-day values to 15 times such values~\citep{2008ProcGeol..119..5P}. 

The simulation results are shown in Fig.~\ref{fig:9}. Both SAT and global ocean temperature increase during the spin-up phase and reach a stable value around 18$^\circ$C and 3.8$^\circ$C, respectively (Fig.~\ref{fig:9}b). The vertical section of the ocean temperature shows that it is well stabilised at all depths (Fig.~\ref{fig:9}c). The upper ocean is much warmer than at the beginning, showing that the initial conditions  should indeed be optimised in order to reduce the gap with respect to the final equilibrium. 
The sea ice extent reaches a constant value of $12\cdot 10^6$~km$^2$ in the north polar region, while it shows larger variability in the south polar 
region near a mean value of $9\cdot 10^6$~km$^2$ (Fig.~\ref{fig:9}d).  

The sensitivity to the tuning parameter, 
quantified by the slope in the plot in Fig.~\ref{fig:6} showing temperature drift against the relative error in $RH$, is of the same order as the sensitivity in the {\tt Run4} series (compare stars and diamonds in Fig.~\ref{fig:6}). We have thus proved that the above method is general and can be applied successfully to other bathymetric configurations.
 
\section{Discussion and conclusions}
\label{section:conclusions}

High-accuracy measurements from satellite programs as CE\-RES (Clouds and the Earth's Radiant Energy System~\citep{1996BAMS...77..853W}) and SORCE (Solar Radiation and Climate Experiment~\citep{2005Anderson}) have constrained the radiation fluxes at TOA with uncertainty range of less than 1 Wm$^{-2}$. The 
resulting imbalance at TOA is in agreement with that determined from changes in ocean heat content that amounts to $0.8\pm 0.2$~Wm$^{-2}$~\citep{2011ACP....1113421H,2013ClDy...40.3107W}. Thus, in the present-day climate, observations point to equal values of TOA and ocean surface budget of the order of 
1~Wm$^{-2}$.
     
Ideally the same imbalance should be obtained in climate models. However, while ensemble averages are in general in agreement with 
these values~\citep{2013ClDy...40.3107W,2014JCli...27.3129T}, there is a huge spread of results in each model of the ensemble, even in pre-industrial control 
simulations (see, for example, Fig.~2 in~\citet{2011RvGeo..49.1001L} for CMIP3 and Fig.~1 in~\citet{2016JCli...29.1639H} for CMIP5). Here, the imbalance should 
ideally be zero~\citep{2011RvGeo..49.1001L} since these simulations represent an estimate of  the unforced quasi-equilibrium climate that evolves under the only 
effect of internal nonlinear dynamics. Moreover, the values of imbalance at TOA and at ocean surface are in general different in climate 
models~\citep{2016JCli...29.1639H}.    

In the control runs presented here, the ocean surface budget is tuned to zero to avoid temperature drift. Therefore 
the imbalance at the ocean surface can be considered as a measure  
of the goodness of the tuning procedure in control runs.  
In our simulations, the TOA budget is different from zero, as shown in Table~\ref{tab:3}. The TOA imbalance can be reduced to lower values by improving the climate code, as in our case where it moved from 2.65 to $-0.55$~Wm$^{-2}$ when friction heating was taken into account in the simulations (compare {\tt Run1} and {\tt Run4} in 
Table~\ref{tab:3}). Therefore, the TOA budget can be considered as a measure of the goodness of parameterisations and coding, since it is related to the presence of energy sources/sinks within the coupled climate system due to imperfect representations of physical processes (such as, for example, friction heating) and/or to numerical diffusion.
Thus, we can confirm the importance of using global metrics, such as TOA and ocean surface imbalance, as first-order diagnostics of the model 
behaviour~\citep{2016JCli...29.1639H}. These two quantities should always be explicitly stated when numerical results are presented. 

A model can perfectly conserve energy between its components at the ocean surface, as the MITgcm~\citep{2008OcMod..24....1C}, and still have energy sources/sinks that affect the TOA budget (due to approximate processes within the model, like numerical or kinetic-energy dissipation),  that are not accounted for and are sometimes called `ghost energy'~\citep{2011RvGeo..49.1001L}.  Within certain models, a correction is applied to eliminate the kinetic energy dissipation by the internal stresses at each time step~\citep{2011ClDy...36.1189P}, while in others, analysis of the results are presented as anomalies with respect to the time-mean nonconservation term~\citep{2016JCli...29.1639H}. We have chosen to explicitly show the values of TOA imbalance since we think that they are very useful to gain insight into the limitations/biases of a given climate model. 

Since we have verified that the net ocean surface heat flux (that is an output of the MITgcm called $TFLUX$) exactly accounts for the temperature drift in the ocean volume, we can state that the energy leaks in MITgcm are mainly outside the ocean, as occurs in the majority of climate models considered in~\citet{2016JCli...29.1639H}. 
This is also confirmed by the fact that the oceanic contribution to the total heat transport, shown in Fig.~\ref{fig:8}b, does not change much for different modelling set-ups, while the total heat transport becomes more closer to the observed value of 6~PW at 30$^\circ$N~\citep{2008JCli...21.2313F} when the correct set-up is implemented 
(compare {\tt Run1} to {\tt Run4}  in Fig.~\ref{fig:8}a).    
 
We have shown that, taken together, the tuning procedure and the parameterisations can give useful insight into limitations of a climate model. An example for such an approach is given by Fig.~\ref{fig:6} where we plotted how the temperature drift $dT/dt$ depends on the tuning parameter $RH$ for different configurations of MITgcm. A strong dependence of the temperature drift on the tuning parameter corresponds to a steep growth (as occurs in {\tt Run3} series, squares), a small change in $RH$ giving rise to a huge temperature drift. This means that the implemented physical process strongly depends on the tuning procedure, with the risk that the resulting control run will be not stable and too strongly dependent on internal variability. 
On the contrary, if small changes in $RH$ produce the same nearly-zero values for the temperature drift, we can expect to obtain good-quality control runs with the considered set-up of the climate code.  
Thus, the analysis of the dependence of the temperature drift on the tuning parameter provides insight on the quality of control runs and their stability.   
This is a general and robust result that can be applied to other climate codes and tuning parameters. 

The importance of developing simple diagnostics to assure the quality of control runs and model verification is even more important in simulations of the 
deep-past where huge uncertainties exist in initial and boundary conditions. From the analysis performed in the previous section, it turns out that the 
tuning procedure should be applied for each different paleogeographic configuration. We stress here the importance of explicitly stating the values of the 
TOA and ocean surface energy imbalance, and how the latter depends on the tuning parameter, each time a new simulation of the deep-past is presented. 
This procedure will indeed allow other researchers to know important aspects on the quality of the numerical results and on the stability of the control run. 
Unfortunately, nowadays these global metrics are almost never mentioned in paleoclimate studies. 
We have seen that a small amount of imbalance in the ocean surface budget can induce large effects 
in SAT, the global ocean temperature or the intensity of the overturning circulation 
(compare {\tt Run1} with {\tt Run2} in Figs.~\ref{fig:1}, \ref{fig:2} and \ref{fig:4}).  This is particularly important in paleoclimate simulations that need to be run for thousands of  years to attain a fully equilibrated coupled atmosphere-ocean state that is not known {\it a priori} (while in present-day simulations we know the characteristics that a control run should satisfy at the end of the simulation). Even a small imbalance, but lasting for a long period of time, can strongly affect the final results. This is why it is important in paleoclimate simulations to correctly tune the parameters from the beginning and use procedures, such as the one illustrated in Fig.~\ref{fig:6}, that allow one to optimise the physical parameterisations and configurations used in the climate code. 
          
In climate simulations of the deep-past there is of course also the need of constructing realistic initial and boundary conditions that can reduce the size of the parameter space. Interdisciplinary effort, with contributions from geologists, physicists, climate modellers, is essential in this respect and intercomparison projects, such as the Paleoclimate Modeling Intercomparison Project~\citep{2011Braconnot,2017GMD....10..889L}, 
are crucial to progress in this field.  Sensitivity studies to the initialisation and to the paleogeography are strongly encouraged. Inclusion of the main variables used in the present study 
(in order to calculate TOA and ocean surface energy imbalance)  are recommended in intercomparison datasets.

\begin{acknowledgements}
The computations were performed at University of Geneva on the Baobab and CLIMDAL3 clusters. 
We thank Jean-Michel Campin, Marjorie Perroud and Martin Beniston for useful discussions, and the MITgcm-support mailing list for valuable advice on the code.
This work was partly supported by CTI 15574.1 PFES-ES. 
\end{acknowledgements}

\bibliographystyle{spbasic} 
\bibliography{paleo} 
    

\end{document}